\documentclass[twoside,twocolumn,9pt]{article}
\usepackage{extsizes}
\usepackage[super,sort&compress,comma]{natbib} 
\usepackage[version=3]{mhchem}
\usepackage[left=1.5cm, right=1.5cm, top=1.785cm, bottom=2.0cm]{geometry}
\usepackage{balance}
\usepackage{mathptmx}
\usepackage{sectsty}
\usepackage{graphicx} 
\usepackage{lastpage}
\usepackage[format=plain,justification=justified,singlelinecheck=false,font={stretch=1.125,small,sf},labelfont=bf,labelsep=space]{caption}
\usepackage{float}
\usepackage{fancyhdr}
\usepackage{fnpos}
\usepackage[english]{babel}
\addto{\captionsenglish}{%
  
}
\usepackage{array}
\usepackage{droidsans}
\usepackage{charter}
\usepackage[T1]{fontenc}
\usepackage[usenames,dvipsnames]{xcolor}
\usepackage{setspace}
\usepackage[compact]{titlesec}
\usepackage{hyperref}
\usepackage{siunitx}
\usepackage{physics}

\definecolor{cream}{RGB}{222,217,201}

\begin{document}

\pagestyle{fancy}
\thispagestyle{plain}
\fancypagestyle{plain}{
\renewcommand{\headrulewidth}{0pt}
}

\makeFNbottom
\makeatletter
\renewcommand\LARGE{\@setfontsize\LARGE{15pt}{17}}
\renewcommand\Large{\@setfontsize\Large{12pt}{14}}
\renewcommand\large{\@setfontsize\large{10pt}{12}}
\renewcommand\footnotesize{\@setfontsize\footnotesize{7pt}{10}}
\makeatother

\renewcommand{\thefootnote}{\fnsymbol{footnote}}
\renewcommand\footnoterule{\vspace*{1pt}%
\color{cream}\hrule width 3.5in height 0.4pt \color{black}\vspace*{5pt}} 
\setcounter{secnumdepth}{5}

\makeatletter 
\renewcommand\@biblabel[1]{#1}            
\renewcommand\@makefntext[1]%
{\noindent\makebox[0pt][r]{\@thefnmark\,}#1}
\makeatother 
\renewcommand{\figurename}{\small{Fig.}~}
\sectionfont{\sffamily\Large}
\subsectionfont{\normalsize}
\subsubsectionfont{\bf}
\setstretch{1.125} 
\setlength{\skip\footins}{0.8cm}
\setlength{\footnotesep}{0.25cm}
\setlength{\jot}{10pt}
\titlespacing*{\section}{0pt}{4pt}{4pt}
\titlespacing*{\subsection}{0pt}{15pt}{1pt}

\fancyfoot{}
\fancyfoot[LO,RE]{\vspace{-7.1pt}\includegraphics[height=9pt]{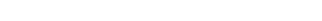}}
\fancyfoot[CO]{\vspace{-7.1pt}\hspace{13.2cm}\includegraphics{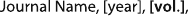}}
\fancyfoot[CE]{\vspace{-7.2pt}\hspace{-14.2cm}\includegraphics{head_foot/RF}}
\fancyfoot[RO]{\footnotesize{\sffamily{1--\pageref{LastPage} ~\textbar  \hspace{2pt}\thepage}}}
\fancyfoot[LE]{\footnotesize{\sffamily{\thepage~\textbar\hspace{3.45cm} 1--\pageref{LastPage}}}}
\fancyhead{}
\renewcommand{\headrulewidth}{0pt} 
\renewcommand{\footrulewidth}{0pt}
\setlength{\arrayrulewidth}{1pt}
\setlength{\columnsep}{6.5mm}
\setlength\bibsep{1pt}

\makeatletter 
\newlength{\figrulesep} 
\setlength{\figrulesep}{0.5\textfloatsep} 

\newcommand{\topfigrule}{\vspace*{-1pt}%
\noindent{\color{cream}\rule[-\figrulesep]{\columnwidth}{1.5pt}} }

\newcommand{\botfigrule}{\vspace*{-2pt}%
\noindent{\color{cream}\rule[\figrulesep]{\columnwidth}{1.5pt}} }

\newcommand{\dblfigrule}{\vspace*{-1pt}%
\noindent{\color{cream}\rule[-\figrulesep]{\textwidth}{1.5pt}} }

\newcommand*\rr[1]{\textcolor{red}{#1}}

\makeatother

\twocolumn[
  \begin{@twocolumnfalse}
{\includegraphics[height=30pt]{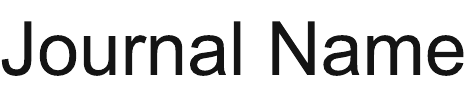}\hfill\raisebox{0pt}[0pt][0pt]{\includegraphics[height=55pt]{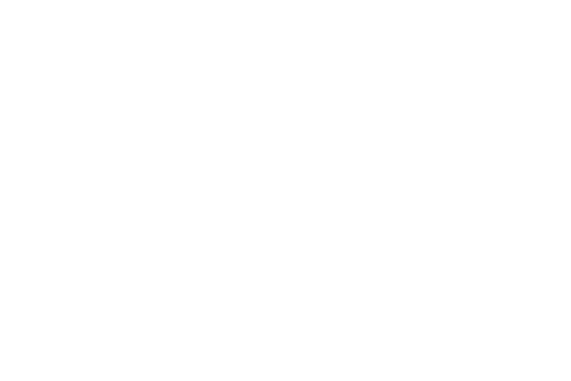}}\\[1ex]
\includegraphics[width=18.5cm]{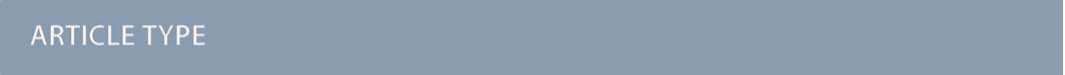}}\par
\vspace{1em}
\sffamily
\begin{tabular}{m{4.5cm} p{13.5cm} }

\includegraphics{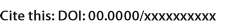} & \noindent\LARGE{\textbf{Doubly resonant enhancement of second-harmonic generation with in-plane phase matching in plasmonic metasurfaces on an AlInP slab waveguide$^\dag$}} \\
\vspace{0.3cm} & \vspace{0.3cm} \\

& \noindent\large{Timo Stolt,\textit{$^{a}$}} Huayu Bai,\textit{$^{a}$} Seyed Ahmad Shahahmadi,\textit{$^{b}$} Jani Oksanen,\textit{$^{b}$} Andriy Shevchenko,\textit{$^{a\ast}$} and Radoslaw Kolkowski$^{a\ast}$ \\

\includegraphics{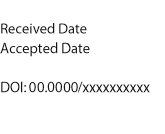} & \noindent\normalsize{Nonlinear metasurfaces have attracted significant interest by offering the possibility to circumvent conventional phase-matching requirements of bulk nonlinear crystals, opening the way to efficient frequency conversion over ultrashort propagation distances. Here, we experimentally demonstrate metasurfaces in which second-harmonic generation is strongly enhanced by in-plane phase matching of high-$Q$ guided-mode resonances. To achieve this enhancement, we use hybrid metasurfaces composed of periodic arrays of gold nanodiscs on a slab waveguide made of aluminum indium phosphide (AlInP) -- a low-loss epitaxial III-V semiconductor material. The metasurfaces are found to exhibit guided-mode resonances with $Q$ factors exceeding 400 and 900 at the visible and near-infrared wavelengths, respectively, demonstrating the unique capabilities of hybrid metal-dielectric structures to support high-$Q$ resonances despite the presence of lossy plasmonic components. Moreover, our frequency and momentum-resolved measurements demonstrate a two-orders-of-magnitude enhancement of second-harmonic generation at doubly resonant conditions. We reproduce the experimental results using numerical simulations, showing that the in-plane phase matching and spatial mode overlap are the main factors governing the enhancement. Our findings demonstrate a promising strategy to realize efficient metasurface-based frequency converters, enabling their potential applications in compact photonic systems.} \\

\end{tabular}

 \end{@twocolumnfalse} \vspace{0.6cm}

  ]

\renewcommand*\rmdefault{bch}\normalfont\upshape
\rmfamily
\section*{}
\vspace{-1cm}


\footnotetext{\textit{$^{a}$~Department of Applied Physics, Aalto University, P.O.Box 13500,  FI-00076 Aalto, Finland, E-mail: andriy.shevchenko@aalto.fi, radoslaw.kolkowski@aalto.fi}}
\footnotetext{\textit{$^{b}$~Engineered Nanosystems Group, Aalto University, P.O.Box 13500, FI-00076 Aalto, Finland}}
\footnotetext{\dag~Supplementary Information available.}





\section{Introduction}

Nonlinear optical processes are the cornerstone of many modern photonic applications, ranging from pulsed and wavelength-tunable lasers to photon pair generation for quantum technologies~\cite{BoydBook2020,GerryQuBook}. A well-known example of such processes is the second-harmonic generation (SHG), in which two photons with the same frequency are annihilated to generate a new photon with a doubled frequency. Consequently, SHG is often used to up-convert laser light from the near-infrared to the visible spectral range. 

SHG arises from a material nonlinearity characterized by the second-order susceptibility $\chi^{(2)}$ inherent in non-centrosymmetric crystals, such as lithium niobate and beta-barium borate. However, in most of these crystals, the intrinsic nonlinear response is extremely weak. To enhance the response to a practical level, special phase-matching techniques are used. They allow minimizing the wavevector mismatch between the interacting light fields of different wavelengths. With the phase matching achieved, the nonlinear frequency conversion shows a high energy conversion efficiency that can be further improved by placing the crystal in an optical resonator~\cite{MillerOPO_PhyLett66}. Unfortunately, conventionally used phase matching can only be realized in birefringent media, excluding isotropic nonlinear materials, such as III-V semiconductors, which can possess high values of $\chi^{(2)}$~\cite{anderson1971wideband,somekh1972phase,fiore1998phase,rivoire2009second}. Therefore, in addition to phase matching using optical anisotropy, other methods have been used, such as quasi-phase matching in periodically poled crystals~\cite{somekh1972phase_,Fejer1992,gordon1993diffusion,pantzas2022continuous}. However, to enable significant device miniaturization, one has to explore alternative approaches, aiming at efficient frequency conversion over much shorter propagation distances. One of such approaches is based on the use of optical metasurfaces~\cite{lee2014giant,LiNat2017,krasnok2018nonlinear,kolkowski2023nonlinear}.

Metasurfaces are two-dimensional arrangements of nanoscale building blocks called meta-atoms, typically fabricated on a transparent substrate~\cite{Meinzer2014,Jahani2016}. Optically resonant meta-atoms, such as plasmonic nanoantennas and Mie-resonant dielectric particles, allow for resonant enhancement of light-matter interactions~\cite{koenderink2015nanophotonics} and efficient nanoscale phase control of optical fields~\cite{Genevet2011}. The local field enhancement provided by such resonant metasurfaces is especially useful for boosting the nonlinear optical response and frequency conversion, which have quadratic and higher-order dependence on the amplitudes of the interacting waves.~\cite{CelebranoNatNano2014,Minkov2019}.

Initially, research on nonlinear metasurfaces focused on metal-based plasmonic nanoantennas~\cite{naturemartti,valev2014nonlinear,kolkowski2015octupolar,Keren-Zur2016}, but later it was extended to include all-dielectric Mie-type resonators~\cite{Liu_res_SHG_diel_NanoLett2016,kivshar2018all,gigli2022}. In dielectric meta-atoms, the ohmic losses of individual resonant meta-atoms are absent, but light trapping and enhancement remain limited. This limitation can be removed by the use of collective resonances in periodic structures, including guided-mode resonances (GMRs)~\cite{quaranta2018recent,Bej2025}, surface lattice resonances (SLRs)~\cite{Hooper2019,Kravets2018}, and waveguide-plasmon polaritons (WPPs)~\cite{christ2003waveguide,rodriguez2012light,zang2023near}. Reducing the radiative losses of such resonances via destructive interference in the far-field can further improve their light-confinement capabilities, giving rise to the so-called dark modes~\cite{rodriguez2011coupling,hakala2017lasing} or quasi-bound states in the continuum (qBIC)~\cite{koshelev2018asymmetric,overvig2020selection}. Dispersive nature of collective resonances (i.e., dependence on both the wavelength and the incidence angle)~\cite{zhou2012} makes them particularly suitable for simultaneous manipulation of the spectral and spatial-frequency components of the fields, governing the design and operation of the so-called nonlocal metasurfaces~\cite{overvig2020multifunctional,overvig2022diffractive,chen2025nonlocal,monticone2025nonlocality,chen2025}. Most importantly, collective resonances can possess high quality ($Q$) factors~\cite{zou2004silver,bin2021ultra} and provide large local field enhancement, which can be used to boost nonlinear optical effects~\cite{kolkowski2016non,MichaeliPRL2017,hooper2018second,koshelev2019,HooperNanoLet2019,Liu2019,anthur2020continuous,zhang2022quasi,zograf2022high,StoltMultires2022,kolkowski2024temporal,Weiss2024,Abir2025}. 

Most of the recently demonstrated high-$Q$ nonlinear metasurfaces are made of virtually lossless dielectric materials that are only weakly nonlinear. On the other hand, highly nonlinear materials, such as III-V semiconductors, exhibit considerable absorption losses in the visible spectral range in which the second harmonic is typically generated, reducing the overall frequency conversion efficiency~\cite{liu2018all}. In this work, we utilize a hybrid metasurface design~\cite{Ju2023,li2024plasmonic,Ansimov2026} that combines a metal and a nonlinear dielectric. Specifically, we use a wide-bandgap zinc-blende-type III-V semiconductor, AlInP, as a nonlinear, optically transparent material having a significant second-order susceptibility~\cite{Shoji2002} and a wide transparency range from the near-infrared down to 500 nm wavelength~\cite{ochoa2018refractive,kolkowski2025low,kolkowski2025nonlinear,kolkowski2025resonant}. A slab of this material, being covered with an array of resonant gold nanoantennas, provides guidance and confinement of light in the form of GMRs, with low intensity at the position of the meta-atoms mitigating the losses via the mode interference~\cite{azzam2018formation,kolkowski2023}. Large on-resonance scattering cross sections of the plasmonic meta-atoms provide efficient coupling of the incident light to GMRs and local enhancement of the confined field. The transparency of AlInP covers both the fundamental and SHG wavelengths, enabling the realization of a condition for doubly resonant SHG enhancement. Being created by epitaxial growth and subsequent wafer bonding, AlInP forms a high-quality crystalline layer with a negligible surface roughness, which minimizes the scattering losses of the guided modes. Due to the above properties, our Au-AlInP metasurfaces support high-$Q$ GMRs, which we use to enhance the intrinsic SHG response of the AlInP layer. Most importantly, we show that the in-plane phase matching and spatial overlap between the fields of GMRs at the fundamental and SHG wavelengths can give rise to a large SHG enhancement, significantly surpassing that obtained away from the doubly resonant condition. Our findings pave the way towards further advancements in the field of resonant nonlinear metasurfaces, bringing them closer to practical applications in compact nonlinear photonic devices.

\begin{figure*}[h!]
    \centering
    \includegraphics[width=\textwidth]{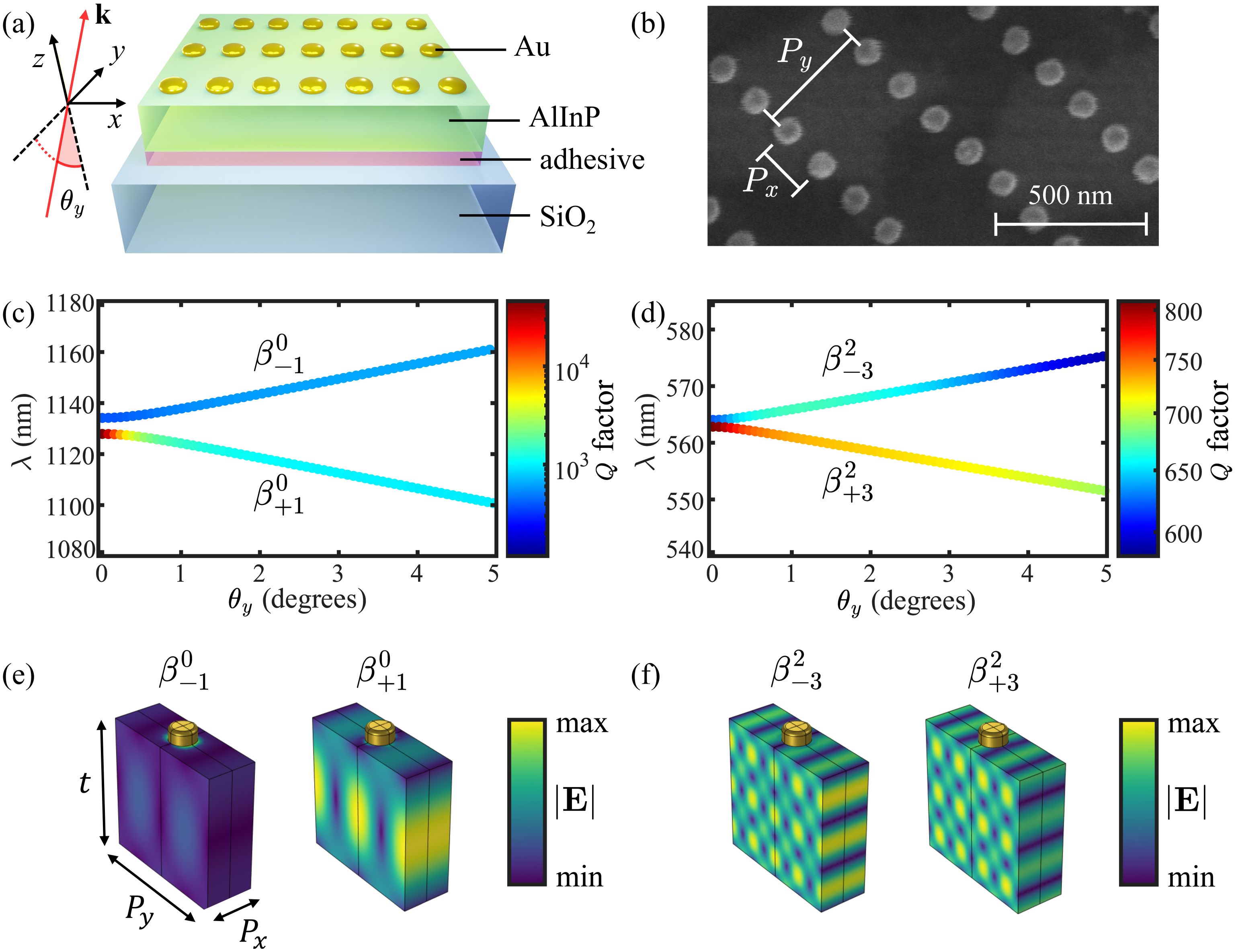}
    \caption{(a) Schematic illustration of a metal-dielectric metasurface, consisting of a periodic array of gold nanodiscs deposited on top of an epitaxial AlInP waveguide, which is wafer-bonded to a glass substrate. The wavevector of incident/outgoing light ($\mathbf{k}$) is illustrated with a red arrow, which shows its orientation in terms of angle $\theta_y$ with respect to the Cartesian axes. (b) Example SEM image of the fabricated sample with periods $P_x=\SI{150}{nm}$ and $P_y=\SI{420}{nm}$. Each nanodisc has a diameter of 80 nm and a thickness of 40 nm, while the thickness of the underlying AlInP layer is around $\SI{400}{nm}$. (c-f) Results of the eigenfrequency analysis performed in COMSOL Multiphysics for different values of the in-plane momentum (different points in the $k$-space) here expressed in terms of the corresponding values of the incidence angle $\theta_y$ in the $yz$-plane. The analysis reveals high-$Q$ transverse magnetic GMRs (with magnetic field $\mathbf{H}$ parallel to $x$) in the (c) infrared and (d) visible spectral ranges. The resonance $Q$ factors are encoded in the colors of the eigenfrequency curves (see the color scales), reaching the highest values at the dark band-edge resonances (qBICs). (e, f) Distributions of the electric field norm $|\mathbf{E}|$ within one unit cell of the structure (with glass substrate made invisible) at $\theta_y=0$ for the GMRs originating from the guided modes with propagation constants $\beta^M_m$ ($\beta^N_n$).}
    \label{fig:sample}
\end{figure*}

\section{Metasurface design and experimental results}
To demonstrate the doubly resonant SHG enhancement by GMRs, we designed and fabricated metasurfaces with a structure shown in Fig.~\ref{fig:sample}a. The metasurfaces are composed of gold nanodiscs arranged into rectangular arrays. Each nanodisc has a height of 40 nm and a diameter of 80 nm. We fabricated and studied several samples with different periods $P_y$ along the $y$-axis. Here, we report on the results obtained with two of these samples: sample A with $P_y=\SI{420}{nm}$ and sample B with $P_y=\SI{360}{nm}$. The period along the $x$-axis is $P_x=\SI{150}{nm}$ in all the samples (see an example scanning electron microscopy image in Fig~\ref{fig:sample}b and additional images in the Electronic Supplementary Information (ESI)$^\dag$). The arrays were fabricated via electron-beam lithography (EBL) on the surface of a 400 nm thick AlInP slab waveguide, which was first grown epitaxially on a gallium arsenide (GaAs) wafer using metal-organic vapor phase epitaxy (MOVPE) and subsequently transferred onto a glass substrate through a wafer-bonding process. We measured the optical constants of AlInP ellipsometrically, confirming its high real part of the refractive index ($n>$ 3 at $\lambda<$ 1000 nm) and wide transparency range (the imaginary part of the refractive index $\kappa$ is $<$ 0.02 for $\lambda>$ 500 nm), making it suitable for efficient confinement of light through low-loss guided modes. Since the surface of AlInP was exposed by chemical wet etching, we also characterized its roughness using atomic force microscopy (AFM), confirming its excellent surface quality (see the ESI$^\dag$ for more details).

The period $P_x$ of the metasurface is small, which makes the arrays dense, increasing the strength of their resonant response. For optical waves propagating along the $x$-axis, the arrays are sub-diffractive, which means that GMRs are primarily tailored by $P_y$ through diffraction along the $y$-axis. In our samples, the (001) crystallographic plane of AlInP coincides with the $xy$-plane of the metasurface, while the [100] crystallographic direction is oriented at $\ang{45}$ with respect to the $x$-axis. In such a configuration, the GMRs originating from transverse-magnetic (TM) guided modes propagating along the $y$-axis can undergo nonlinear interactions through nonzero tensor elements $\chi^{(2)}_{yyz}$, $\chi^{(2)}_{yzy}$, and $\chi^{(2)}_{zyy}$ (see ESI$^\dag$). 

Excitation of GMRs is governed by their phase matching to the incident wave as
\begin{equation}
    \beta^M_m(\lambda_1,\theta_{y1})=k_y(\lambda_1,\theta_{y1}),
    \label{eq:beta1}
\end{equation}
where integers $M$ and $m$ indicate the orders of the guided mode and diffraction, respectively, $\lambda_1$ and $\theta_{y1}$ are the wavelength of the incident beam and its incidence angle in the $yz$-plane, respectively, $\beta_m^M(\lambda_1,\theta_{y1})$ is the propagation constant of the guided Bloch mode, and $k_y(\lambda_1,\theta_{y1})$ is the $y$-component of the incident wavevector. The incident light is a collimated beam that initially propagates in air. Its in-plane wavevector component can be written as $k_y(\lambda_1,\theta_{y1})=\sin(\theta_{y1})2\pi/\lambda_1$. 

Under the doubly resonant condition, the far-field radiation from a second-harmonic GMR is governed by
\begin{equation}
\beta^N_n(\lambda_2,\theta_{y2})=k_y(\lambda_2,\theta_{y2}),
    \label{eq:beta2}
\end{equation}
where $\beta_n^N(\lambda_2,\theta_{y2})$ is the propagation constant of the second-harmonic Bloch mode with orders $N$ and $n$. Assuming that the SHG photons are emitted at the same angle at which the pump beam is incident on the other side of the metasurface ($\theta_{y1}=\theta_{y2}=\theta_{y}$) and using the relation $\lambda_2=\lambda_1/2$, we obtain
\begin{equation}
    k_y(\lambda_2,\theta_{y})=\sin(\theta_{y})2\pi/\lambda_2=2\sin(\theta_{y})2\pi/\lambda_1=2k_y(\lambda_1,\theta_{y}).
\end{equation}
This means that, under the doubly resonant condition, the phase matching requirement
\begin{equation}
    \Delta\beta =\beta^N_n(\lambda_2,\theta_y)-2\beta^M_m(\lambda_1,\theta_y)=k_y(\lambda_2,\theta_{y})-2k_y(\lambda_1,\theta_{y})=0
\end{equation}
is automatically satisfied.

\begin{figure*}[h!]
    \centering
    \includegraphics[width=\textwidth]{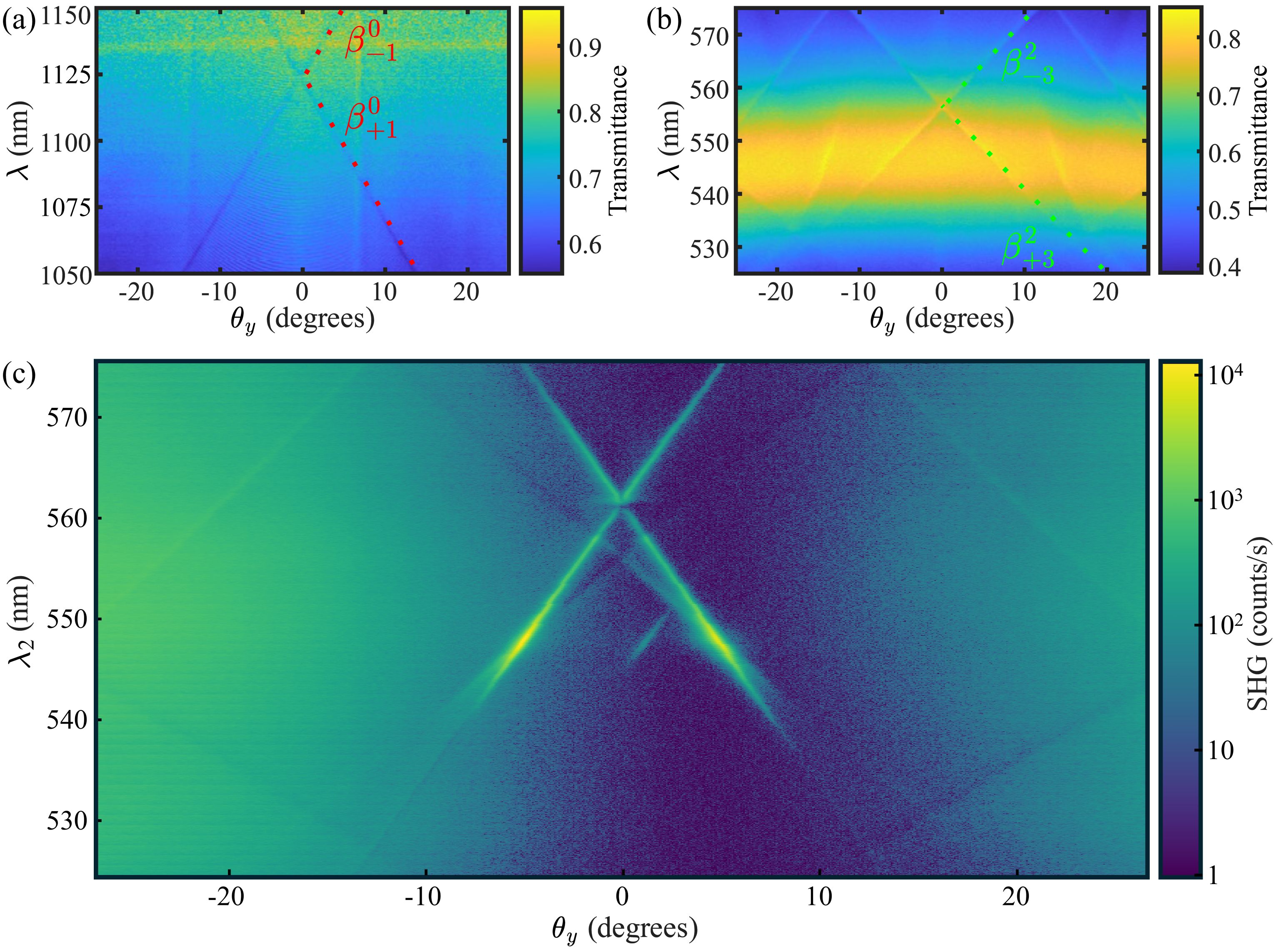}
    \caption{(a, b) Transmission of collimated TM-polarized light from a broadband incoherent source, measured as a function of the wavelength and incidence angle $\theta_y$ for sample A ($P_y=\SI{420}{nm}$) in the (a) infrared and (b) visible spectral ranges. Red and green dotted lines highlight the GMRs that contribute to the measured SHG enhancement presented in (c).}
    \label{fig:Py420_T}
\end{figure*}

The metasurfaces are designed to enable doubly resonant SHG enhancement for fundamental wavelengths ($\lambda_1$) around 1100 nm and SH wavelengths ($\lambda_2$) around 550 nm. In sample A ($P_y=\SI{420}{nm}$), the eigenfrequency analysis reveals high-$Q$ GMRs at near normal incidence, associated with the Bloch modes marked by their propagation constants $\beta^0_{\pm1}$ and $\beta^2_{\pm3}$ in Fig.~\ref{fig:sample}c and d (in the further text, we will refer to GMRs associated with the Bloch modes with propagation constants $\beta^M_m$ briefly as ``$\beta^M_m$ GMRs''). Their dispersion curves for the positive and negative diffraction orders anti-cross at $\theta_y=0$ in both wavelength ranges (near 1130 nm and 562 nm,\footnote{Due to the dispersion of optical constants, the exact values of the resonant wavelengths obtained by the eigenfrequency analysis depend slightly on the simulation settings, e.g., the frequency around which the eigenmodes are searched for.} respectively), giving rise to bright and dark band-edge resonances. In the near-infrared, the bright resonance corresponds to a GMR standing wave with nodes located at the positions of the nanodiscs. It is characterized by a high magnitude of the $y$-component of the electric field at the nanodiscs (see the left plot in Fig.~\ref{fig:sample}e). In the case of the dark resonance (qBIC), the evanescent field is primarily polarized in the $z$-direction, avoiding strong interaction with the nanodiscs and minimizing the ohmic and radiation losses simultaneously~\cite{kolkowski2023}. In the visible, the situation is reversed, i.e., the lower-$Q$ bright mode corresponds to the nanodisc at the anti-node of the GMR. This difference can be attributed to the short period of the standing wave, that becomes comparable to the lateral size of the nanodisc, and the fact that the gold nanodisc is a less efficient scatterer at shorter wavelengths. The eigenfrequency analysis shows that the dark GMRs exhibit very high $Q$ factors near $\theta_y=0$, on the order of $10^4$ and $10^3$ in the spectral ranges of the fundamental and SH waves, respectively. 

\begin{figure*}[h!]
    \centering
    \includegraphics[width=\textwidth]{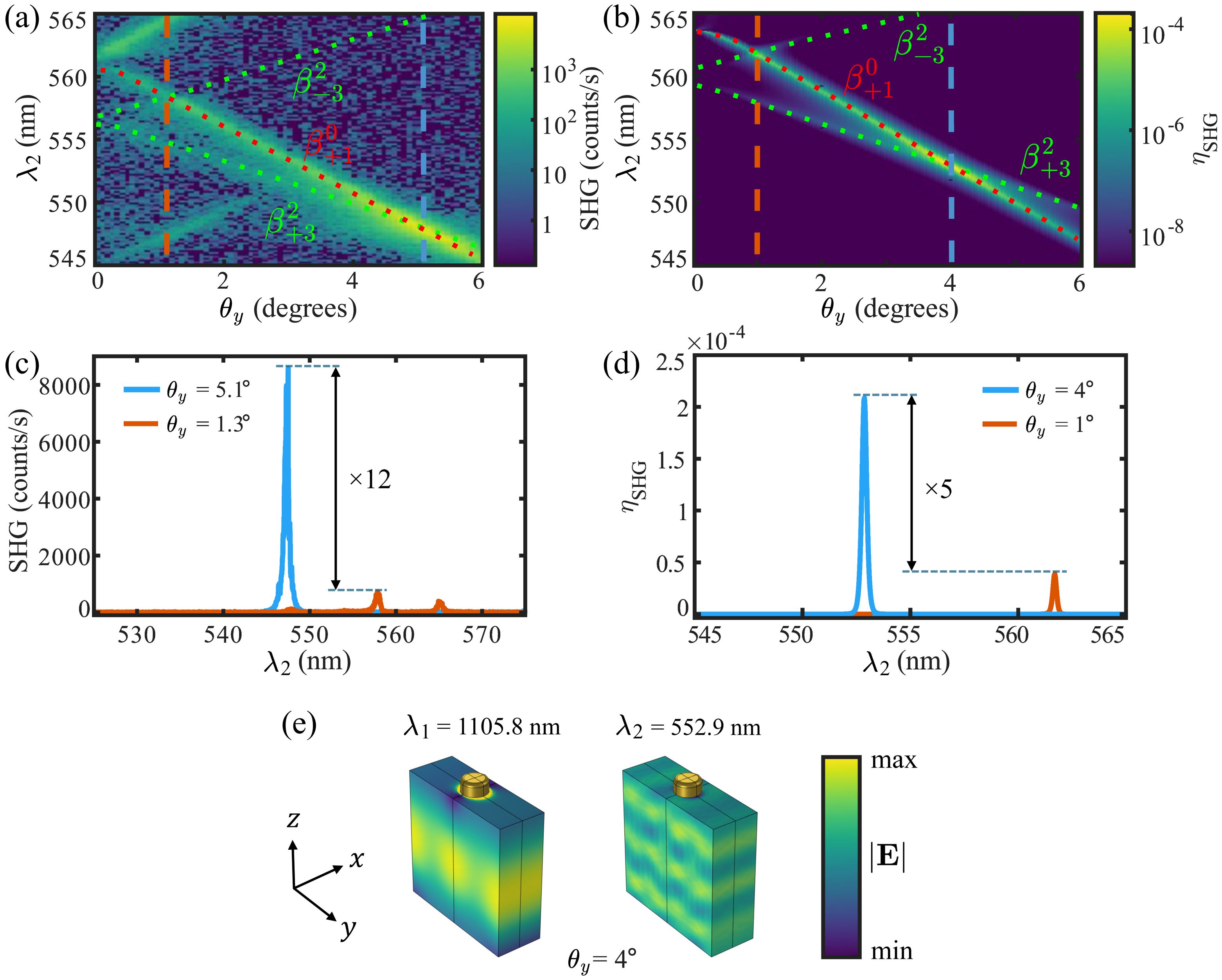}
    \caption{(a) Measured and (b) simulated dependence of the SHG on the wavelength ($\lambda_2$) and incidence angle ($\theta_y$, equivalent to the emission angle) for sample A ($P_y=\SI{420}{nm}$). The red dotted lines indicate the frequency-doubled dispersion of the $\beta_{+1}^0$ GMR, while the green dotted lines indicate the dispersion of the $\beta_{\pm3}^2$ GMRs. The orange and blue dashed lines highlight the doubly resonant conditions. Note that both SHG maps are plotted on a logarithmic color scale. The data and labels are presented separately in the ESI$^\dag$ to show some of the modes more clearly. (c) Measured and (d) simulated SHG spectra at the two doubly resonant conditions, indicated by the vertical dashed lines in (a) and (b). The spectra are plotted on a linear scale. (e) Simulated distributions of the electric-field norm $\abs{\mathbf{E}}$ under the doubly resonant condition at $\theta_y=\ang{4}$, pump wavelength $\lambda_1=\SI{1105.8}{nm}$, and SHG wavelength $\lambda_2=\SI{552.9}{nm}$. }
    \label{fig:SHG_Py420}
\end{figure*}

Experimentally, GMRs in question are visible as narrow lines in the transmission of $y$-polarized light (see Fig.~\ref{fig:Py420_T}a,b), which we measured as a function of the wavelength and incidence angle $\theta_y$ (in Fig.~\ref{fig:Py420_T}a/b, the GMRs reduce/increase the transmittance contributing more to the reflected/transmitted light). The measurements were performed using a spectrally-resolved back-focal-plane imaging setup, with a quartz tungsten-halogen lamp serving as the light source, a galvo mirror used for scanning the $k$-space (in combination with an aspheric lens and a high-NA objective), and an imaging spectrometer utilized for signal detection (with the transmitted light collected by another high-NA objective, having its back focal plane imaged on the spectrometer's entrance slit; see ESI$^\dag$). These transmission measurements yield GMR intersections at $\theta_y=0$ at the wavelengths of 1125 nm and 556 nm for the fundamental and SH waves, respectively, which are close to the simulated values. The anti-crossing is not clearly resolved in these measurements, but the spectral widths of the GMRs can be coarsely estimated at $\Delta\lambda\approx\SI{1}{nm}$, corresponding to the $Q$ factors of about 1000 and 500 in the infrared and visible spectral ranges, respectively. In addition to TM-polarized GMRs, our samples also exhibit transverse-electric (TE) GMRs, but we have found that they do not play a significant role in the SHG enhancement, and therefore, we do not discuss them here. Their transmission spectra are shown in the ESI$^\dag$.

Having established that sample A exhibits high-$Q$ GMRs in the appropriate spectral regions, we subsequently measured its SHG emission as a function of the SH wavelength and emission angle (see Fig.~\ref{fig:Py420_T}c). We illuminated the sample from one side with a $y$-polarized beam of a tunable femtosecond laser (see ESI$^\dag$), scanning the incidence angle (as in the linear transmission measurements) and collecting the SHG signal with an objective located on the other side of the sample. The SHG signal was passed through a dichroic mirror and a set of short-pass filters blocking the fundamental beam. To probe the GMRs, we tuned the fundamental wavelength $\lambda_1$ in the range $\SI{1050}{}$--$\SI{1150}{nm}$, corresponding to the SH wavelength $\lambda_2$ in the range $\SI{525}{}$--$\SI{575}{nm}$. The obtained two-dimensional map of the measured SHG signal is presented in Fig.~\ref{fig:Py420_T}c, showing clear enhancement at the GMRs. The region of the doubly resonant condition is magnified in Fig.~\ref{fig:SHG_Py420}a.

To accurately identify the GMRs participating in the experimentally observed SHG enhancement, we modeled the SHG process via two-step full-wave frequency-domain scattering simulations in COMSOL Multiphysics (see ESI$^\dag$). The calculated far-field SHG power, expressed as the conversion efficiency $\eta_{\text{SHG}}$, is presented in Fig.~\ref{fig:SHG_Py420}b. The simulated SHG map is in good agreement with the corresponding experimental results in Fig.~\ref{fig:SHG_Py420}a, except for a slight spectral shift of the GMR peaks. Moreover, the experimental data contain some additional features that are not reproduced by the simulations -- these may result from the SHG processes that do not conserve the in-plane momentum (e.g., due to scattering by non-periodic defects or reflection of the propagating Bloch modes from one of the edges of the array). Both experimental and theoretical maps show the highest SHG enhancement under the doubly resonant condition, where the frequency-doubled dispersion line of the $\beta^{0}_{+1}$ GMR at the pump wavelength (red dots) intersects with the dispersion line of the co-propagating $\beta^2_{+3}$ GMR at the SHG wavelength (green dots). This intersection occurs at $\theta_y=\ang{5.1}$ in the experimental results and $\theta_y=\ang{4}$ in the simulations (the angles are indicated by the vertical blue dashed lines in Fig.~\ref{fig:SHG_Py420}a and b). Another doubly resonant condition occurs at the intersection of the $\beta^{0}_{+1}$ GMR with the counter-propagating $\beta^2_{-3}$ GMR. This intersection is marked by the dashed orange lines at $\theta_y=\ang{1.3}$/$\ang{1}$ in the experiment/simulation. The SHG enhancement by the $\beta^2_{-3}$ mode is relatively weak. To show it more clearly, we present another version of Fig.~\ref{fig:SHG_Py420}a and b in the ESI$^\dag$, with the data and labels displayed separately. The measured and calculated SHG spectra in Fig.~\ref{fig:SHG_Py420}c and d show the vertical crosscuts through the two-dimensional maps along the vertical dashed lines in Fig.~\ref{fig:SHG_Py420}a and b. In Fig.~\ref{fig:SHG_Py420}e, we plot the simulated electric field distributions at the pump and SHG wavelengths at $\theta_y=\ang{4}$, which confirm that the SHG enhancement indeed involves the $\beta^{0}_{+1}$ and $\beta^{2}_{+3}$ GMRs.

In doubly resonant structures, the total enhancement of the far-field SHG power $\mathcal{P}_{\text{SHG}}$ depends on the local field enhancement factors $L_1$ and $L_2$ at the fundamental and SH wavelengths~\cite{chen1983surface,maierplasmonics} and the spatial overlap $\Gamma$ of the fundamental and SH fields~\cite{rodriguez2007chi,Millerrule_Obrien_natmate,lin2016cavity,minkov2019doubly,zanotti2021doubly,ge2023doubly,li2026optimizing}
\begin{equation}
    \mathcal{P}_{\text{SHG}}\propto |L_1^2|^2|L_2|^2|\Gamma|^2.
    \label{eq:E_far}
\end{equation}
Note that $L_1$ and $L_2$ are defined for specific excitation conditions, which means that they take into account the efficiency of coupling between the GMRs and the far-field radiation (i.e., excitation by the pump in $L_1$ and SHG emission in $L_2$).  Equation~(\ref{eq:E_far}) shows that the SHG far-field power scales as the fourth power of the local field enhancement at the pump wavelength ($L_1$), which explains the strong contribution of the $\beta^{0}_{+1}$ mode. In contrast, the local field enhancement at the SHG wavelength ($L_2$) contributes only with its second power, which makes the contributions of the $\beta^{2}_{\pm3}$ modes significantly less pronounced in the SHG maps. 

The most interesting observation is that the doubly resonant condition at the $\beta^{0}_{+1}$-$\beta^{2}_{+3}$ curve intersection gives rise to a much stronger SHG enhancement than at the $\beta^{0}_{+1}$-$\beta^{2}_{-3}$ intersection in both the experimental and theoretical results. In the experimental data (Fig.~\ref{fig:SHG_Py420}c), the maximum enhancement at $\theta_y=\ang{5.1}$ ($\beta^{0}_{+1}$-$\beta^{2}_{+3}$ intersection) is about 12 times larger than the maximum enhancement at $\theta_y=\ang{1.3}$ ($\beta^{0}_{+1}$-$\beta^{2}_{-3}$ intersection). The difference in the enhancement at the two doubly resonant conditions in the theoretical curves in Fig.~\ref{fig:SHG_Py420}d is slightly smaller (a factor of 5, compared to 12 in the experiment), but it is still remarkable.

\begin{figure*}[h]
    \centering
    \includegraphics[width=\textwidth]{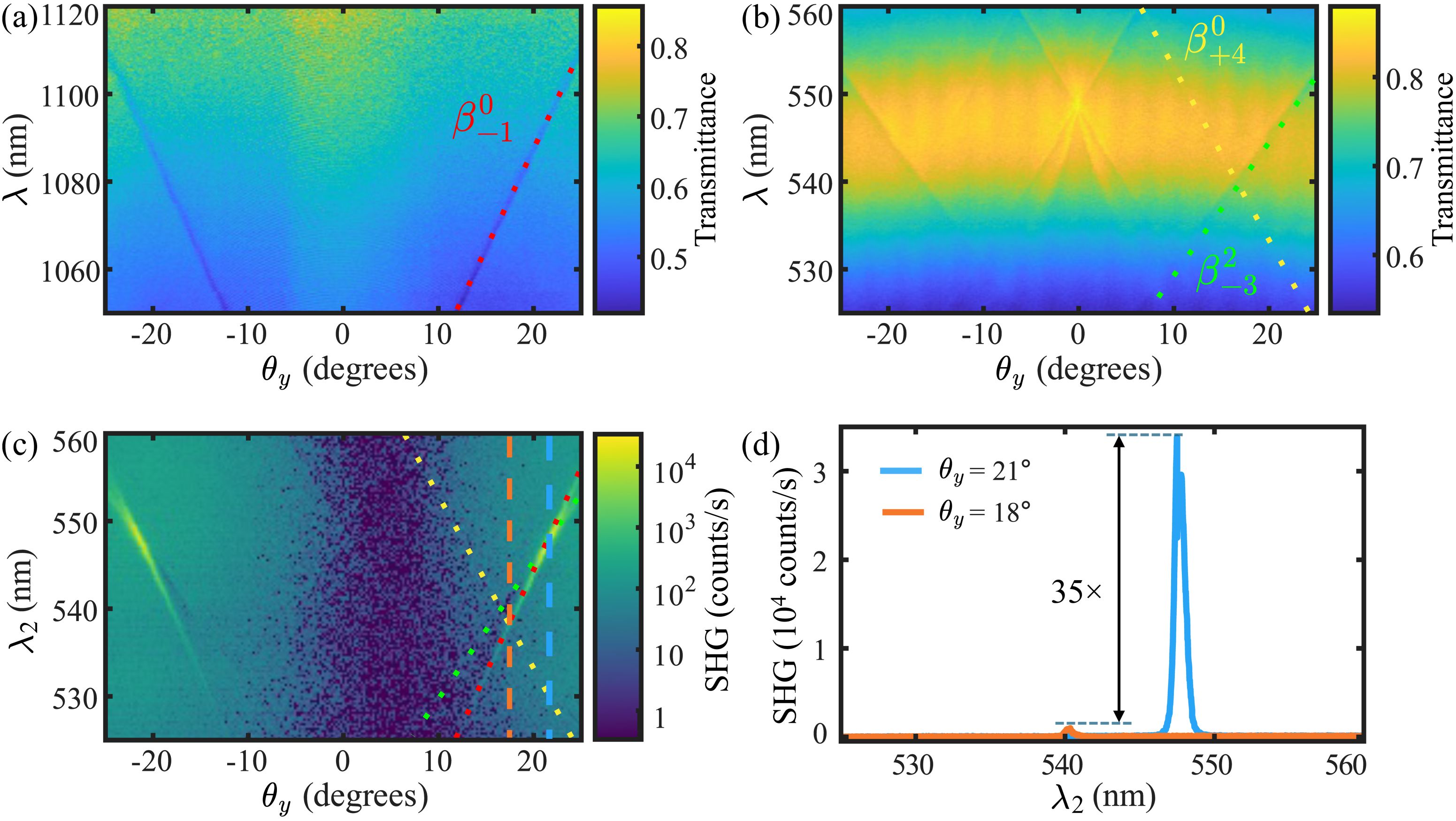}
    \caption{Experimental results for sample B ($P_y = $ 360 nm). (a, b) Broadband transmittance measured as a function of incidence angle $\theta_y$ and wavelength in the (a) pump and (b) SHG spectral ranges. Red, green, and yellow dotted lines indicate the GMRs of interest. (c) SHG signal measured as a function of emission angle $\theta_y$ and SHG wavelength. Dotted lines follow the dispersion of the frequency-doubled GMR at the pump wavelengths (red) and higher-order GMRs at the SHG wavelengths (green and yellow). The angles at which doubly resonant conditions occur ($\ang{18}$ and $\ang{21}$) are marked with the orange and blue vertical dashed lines. (d) SHG spectra corresponding to the vertical crosscuts along the dashed lines in (c). The doubly resonant case at $\theta_y=\ang{21}$ gives rise to 390-fold SHG enhancement with respect to the off-resonant SHG, which is 35 times larger than for the doubly resonant case at $\theta_y=\ang{18}$.}
    \label{fig:Py360_results}
\end{figure*}

The above observations can be explained by investigating the mode overlap factor $\Gamma$ (see Eq.~(\ref{eq:E_far})), which in the general case can be expressed as~\cite{rodriguez2007chi,Millerrule_Obrien_natmate,lin2016cavity,minkov2019doubly,zanotti2021doubly,ge2023doubly,li2026optimizing}
\begin{equation}
    \Gamma=\lambda_1^{3/2}\frac{\int d\mathbf{r}\sum_{ijk}\hat{\chi}^{(2)}_{ijk}E_{i2}^{*}E_{j1}E_{k1}}{[\int d\mathbf{r}\epsilon_1|\mathbf{E}_1|^2][\int d\mathbf{r}\epsilon_2|\mathbf{E}_2|^2]^{1/2}},
    \label{eq:Gamma}
\end{equation}
where $\hat{\chi}^{(2)}_{ijk}$ are the normalized dimensionless elements of the second-order susceptibility tensor, $E_{i1}$ and $E_{i2}$ are the Cartesian electric field components of the fundamental and SH modes, respectively, $\epsilon_1$ and $\epsilon_2$ are the relative electric permittivities of the nonlinear material at the fundamental and SH wavelengths, respectively, and the integrals are calculated over the volume of the nonlinear material. In our case, we can safely assume that the measured SHG originates primarily from the AlInP layer, within which the pump and SHG fields propagate and interact. Plasmonic nanoparticles can also contribute to the SHG signal through the bulk and surface $\chi^{(2)}$ nonlinearity of gold~\cite{bachelier2010origin,kolkowski2015octupolar} which can be enhanced by localized plasmon resonances and surface defects that break the inversion symmetry of the nanoparticles. However, this SHG contribution would be observed at all incidence and emission angles, including the normal direction. Since we do not observe such an emission, the contribution to the SHG of the plasmonic nanoparticles can be neglected. 
Considering the $\chi^{(2)}$-tensor symmetry of AlInP and the illumination conditions used in our experiment, the integrand in the numerator of Eq.~(\ref{eq:Gamma}) can be expressed as 
\begin{equation}
    \sum_{ijk}\hat{\chi}^{(2)}_{ijk}E_{i2}^{*}E_{j1}E_{k1}=E_{z2}^{*}E_{y1}^2+2E_{y2}^{*}E_{y1}E_{z1},
    \label{eq:numerator}
\end{equation}
where $\hat{\chi}^{(2)}_{zyy}=\hat{\chi}^{(2)}_{yyz}=\hat{\chi}^{(2)}_{yzy}=1$. To calculate the mode overlap factor $\Gamma$ for the $\beta^{0}_{+1}$ and $\beta^{2}_{+3}$ modes at the doubly resonant condition, we calculated their electric field profiles using the eigenfrequency analysis at the $k_y$ value that corresponds to $\theta_y=\ang{4}$. Next, we calculated $\Gamma$ for the $\beta^{0}_{+1}$ and $\beta^{2}_{-3}$ modes in the same way, setting $k_y$ to that corresponding to $\theta_y=\ang{1}$. To make our comparison independent of the normalization method, we evaluated the ratio 
\begin{equation}\rho_{\text{A}}=\frac{|\Gamma(\beta^{0}_{+1},\beta^{2}_{+3})|}{|\Gamma(\beta^{0}_{+1},\beta^{2}_{-3})|},
\end{equation} 
which is found to be equal to 5.1. This means that the overlap factor for the co-propagating $\beta^{0}_{+1}$ and $\beta^{2}_{+3}$ modes is five times larger than for the counter-propagating $\beta^{0}_{+1}$ and $\beta^{2}_{-3}$ modes, which should lead to 25 times more efficient SHG. However, as can be seen in Fig.~\ref{fig:sample}, the $\beta^{2}_{+3}$ mode has a higher $Q$ factor (i.e., smaller radiative loss) than the $\beta^{2}_{-3}$ mode. This means that the SHG radiation from the $\beta^{2}_{+3}$ mode is weaker than from the $\beta^{2}_{-3}$ mode, which is why the SHG emission is enhanced by the mode overlap only 5 times in the simulations and 12 times in the experiment.

In the experiment, an additional factor that contributes to the observed SHG efficiency is the spatiotemporal walk-off of the fundamental and SH GMRs under the femtosecond pulsed excitation. This walk-off depends on the relative group velocity, i.e., the difference between the slopes of the dispersion curves of the frequency-doubled pump GMR and the SH GMR. Specifically, the walk-off may deteriorate the observed doubly resonant SHG enhancement for counter-propagating modes (such as $\beta^{0}_{+1}$ and $\beta^{2}_{-3}$) compared to the co-propagating modes (such as $\beta^{0}_{+1}$ and $\beta^{2}_{+3}$), which is consistent with our experimental observations. However, such spatiotemporal effects are not included in our simulations, which are performed in the frequency domain and take into account only one unit cell (i.e., the system is assumed to be infinitely periodic in space and time). Furthermore, these effects are minimized in our experiments, as we illuminate the metasurfaces by plane waves that are uniform over a large area.

To estimate the $Q$ factors of the GMRs, we analyzed the spectral widths of the peaks at $\lambda_2$ in Fig.~\ref{fig:SHG_Py420}a and used the relation $Q\approx\lambda/\Delta\lambda$, where $\Delta\lambda$ is the full width at half maximum (FWHM). The analysis is presented in the ESI$^\dag$. For $\theta_y$ in the range $1.5$--$\ang{2.5}$, the frequency-doubled GMR dispersion of the $\beta^0_{+1}$ mode is well isolated from the neighboring SH GMRs. Fitting the averaged GMR peak with a Lorentzian curve gives $\Delta\lambda_2\approx$ 1.23 nm at $\lambda_2$ in the range 554.5--557 nm. This gives $Q=2\lambda_2/\Delta\lambda_2\approx900$ for the $\beta^0_{+1}$ mode at the pump wavelength $\lambda_1\approx$ 1112 nm. For $\theta_y$ in the range $0.5$--$\ang{1.5}$, the isolated SHG via the $\beta^2_{-3}$ mode has an average $\Delta\lambda_2$ equal to 1.29 nm at $\lambda_2$ in the range 553.5--555.5 nm. This yields $Q=\lambda_2/\Delta\lambda_2\approx430$ for the $\beta^2_{-3}$ mode at the SH wavelength.


\begin{table*}
\small
  \caption{\ Comparison of the SHG enhancement ratios and the mode overlap ratios at the doubly resonant conditions}
  \label{tab:tab1}
  \begin{tabular*}{\textwidth}{@{\extracolsep{\fill}}lllllllll}
    \hline
    Sample & $P_y$ (nm) & GMRs & $\lambda_2$ range (nm) & $\theta_y$ range ($^\circ$) & \multicolumn{2}{l}{SHG enhancement ratio} & Ratio between the \\
    \, & \, & \, & \, & \, & * measured  & * simulated & mode overlap factors \\
    \hline
    A & 420 & $\beta^{0}_{+1}$-$\beta^{2}_{+3}$ vs $\beta^{0}_{+1}$-$\beta^{2}_{-3}$ & 545--565 & 1--6 & 12 & 5 & $\rho_{\text{A}}=5.1$ \\
    B & 360 & $\beta^{0}_{-1}$-$\beta^{2}_{-3}$ vs $\beta^{0}_{-1}$-$\beta^{0}_{+4}$ & 540--555  & 17--22 & 35 & 200 & $\rho_{\text{B}}=33.2$\\
    \hline
  \end{tabular*}
\end{table*}

Next, we turn to sample B with $P_y=\SI{360}{nm}$, in which doubly resonant conditions occur at much larger incidence angles ($\theta_y$ around $\ang{20}$). At such angles, the off-resonant SHG signal from the AlInP layer becomes sufficiently bright to be used as a reference in evaluating the resonant enhancement. The measured transmission and SHG spectra are presented in Fig.~\ref{fig:Py360_results}. The transmission measurements in the pump-wavelength range reveal only one GMR, namely $\beta^0_{-1}$ (red dotted line in Fig.~\ref{fig:Py360_results}a), while at the SHG wavelengths, we find two GMRs at similar incidence angles, $\beta^0_{+4}$ and $\beta^2_{-3}$ (yellow and green dotted lines in Fig.~\ref{fig:Py360_results}b, respectively). The signatures of these GMRs can be seen in the measured SHG enhancement map (see Fig.~\ref{fig:Py360_results}c). The doubly resonant conditions are found at $\theta_y=\ang{18}$ for the $\beta^0_{-1}$-$\beta^0_{+4}$ mode intersection and $\theta_y=\ang{21}$ for the intersection of the $\beta^0_{-1}$ and $\beta^2_{-3}$ modes, marked with the orange and blue vertical dashed lines, respectively. Figure~\ref{fig:Py360_results}d shows the corresponding spectral crosscuts. The maximum SHG enhancement achieved at the two doubly resonant conditions differs by a factor of 35, which is much larger than for sample A. Furthermore, under the phase-matched doubly resonant condition at $\theta_y=\ang{21}$, the overall SHG enhancement with respect to the off-resonant SHG response reaches a staggering factor of 390. The corresponding theoretical results for sample B are presented in the ESI$^\dag$. Remarkably, the simulations predict that the SHG enhancement at the doubly resonant condition achieved with the $\beta^0_{-1}$ and $\beta^2_{-3}$ modes is 200 times larger compared to the case dealing with the $\beta^0_{-1}$ and $\beta^0_{+4}$ modes. We also calculated the ratio between the mode overlap factors
\begin{equation}\rho_{\text{B}}=\frac{|\Gamma(\beta^0_{-1},\beta^2_{-3})|}{|\Gamma(\beta^0_{-1},\beta^0_{+4})|},
\end{equation}
where $\Gamma(\beta^0_{-1},\beta^2_{-3})$ and $\Gamma(\beta^0_{-1},\beta^0_{+4})$ are calculated at $\theta_y=\ang{21.2}$ and $\ang{17.2}$, respectively. We find $\rho_{\text{B}}$ to be equal to 33.2, which should give rise to a $\sim$1000 times greater SHG enhancement at the $\beta^0_{-1}$-$\beta^2_{-3}$ intersection compared to that at the $\beta^0_{-1}$-$\beta^0_{+4}$ intersection. The observed difference in the SHG enhancement is smaller (which, again, may be attributed to the different radiation efficiencies of the SH modes), but it agrees qualitatively with the corresponding ratio of the mode overlap factors. The results for both samples are summarized in Table~\ref{tab:tab1}, showing that the mode overlap correlates very strongly with the SHG enhancement in the studied systems.

\section{Discussion}

The above results demonstrate that the efficiency of nonlinear optical effects in nonlocal metasurfaces can be greatly improved by simultaneous optimization of the phase matching and spatial overlap between guided-mode resonances at the wavelengths of the interacting waves. In more conventional ``local'' metasurfaces, nonlinear effects occur locally within the individual unit cells, so that their efficiency is primarily governed by the local-field enhancement and mode overlap, whereas the in-plane phase matching of the waves is not relevant. In our case, the nonlinear response originates from the resonant modes propagating in the plane of the metasurface and interacting with many unit cells simultaneously. Therefore, phase matching of the fundamental and second-harmonic GMR fields is crucial, while their spatial overlap determines further improvement of the SHG efficiency.

In addition to the doubly resonant enhancement of SHG, our results demonstrate a unique capability of hybrid plasmonic-dielectric metasurfaces to exhibit high-$Q$ GMRs. Typically, achieving $Q>100$ is challenging with metasurfaces relying on plasmonic nanostructures, especially in the visible spectral range. However, in our design, the light is confined mainly in the lossless AlInP waveguide, avoiding the ohmic losses in gold nanodiscs. At the same time, the periodically distributed nanodiscs enable coupling between GMRs that results in the formation of dark modes (qBICs) with suppressed radiation loss in the anti-crossing regions (see Fig.~\ref{fig:sample}c and d). Simultaneously, the electric field distributions of these modes have minima that coincide with the locations of the nanodiscs, leading to further reduction of the ohmic losses (see Fig.~\ref{fig:sample}e and f). The above mechanisms allow us to achieve very high $Q$ factors even in the visible spectral range ($Q\approx430$ at $\lambda\approx$ 555 nm). The dispersive nature of GMRs allows the metasurfaces to be designed towards doubly resonant conditions at various wavelengths and incidence angles, which can be used to realize spectrally and directionally tunable frequency converters.

Last but not least, our results showcase AlInP as a promising material for nonlinear photonic devices. As a wide-bandgap III-V semiconductor possessing noncentrosymmetric zinc-blende crystal structure, AlInP has a considerable intrinsic nonlinearity ($d_{36}=\SI{36}{pm/V}$), while it is transparent over a wide spectral range (practically no absorption down to $\lambda=\SI{500}{nm}$) and has a relatively high refractive index ($n>3$) across the visible and near-infrared regions (see the ESI$^\dag$ for more details). Importantly, AlInP is lattice-matched to GaAs, which allows AlInP thin films to be epitaxially grown on commercial GaAs wafers and subsequently transferred to transparent substrates via wafer bonding. The above advantages of AlInP have so far been largely overlooked by the photonics community, which we hope will change after showcasing its potential in this work.

The specific approaches that we used to realize high-$Q$ resonances and enhanced nonlinear optical effects can be adapted to a variety of different systems and material platforms. For example, gold nanodiscs could be replaced by other types of scatterers, such as Mie-resonant nanostructures or even air holes. Instead of TM modes, one can utilize TE-polarized GMRs to boost nonlinear effects  excited at normal incidence, while the waveguide could be made of another low-loss nonlinear material, such as lithium niobate~\cite{Ju2023,li2024plasmonic,Ansimov2026}. We also foresee that diverse multi-resonant and phase-matching scenarios could be realized to enhance other nonlinear frequency conversion processes, including sum-frequency generation and spontaneous parametric down-conversion. Such demonstrations could pave the way towards miniaturized counterparts of traditional nonlinear crystals used in wavelength-tunable lasers and photon-pair sources. Finally, we have demonstrated metasurfaces operating as free-space optical elements, but we believe that our design could also be adapted to photonic integrated circuits, opening new possibilities in nonlinear guided-wave manipulation.

\section*{Conclusions}

In this work, we used hybrid plasmonic-dielectric metasurfaces to achieve doubly resonant enhancement of second-harmonic generation by high-$Q$ guided-mode resonances, demonstrating that their phase matching and mode overlap play crucial roles in optimizing the nonlinear frequency conversion efficiency. In particular, we have demonstrated that optimizing the mode overlap between the phase-matched GMRs allows the SHG enhancement to be improved by a large factor (up to 35 in the experiments and up to 200 in the simulations), which allowed us to reach a total enhancement of 390 compared to the off-resonant SHG reference. These remarkable observations were enabled by the dispersive nature and high $Q$ factors of the GMRs, which we achieved at both visible and near-infrared wavelengths ($Q\approx$ 430 at $\lambda\approx$ 555 nm and $Q\approx$ 900 at $\lambda\approx$ 1112 nm) despite the presence of lossy plasmonic nanoparticles, owing to the unique capabilities of hybrid systems to redistribute optical fields in a way avoiding losses. In addition, our results demonstrated AlInP as a promising III-V material for low-loss nonlinear photonics. Our approaches can be generalized to other material platforms, different nonlinear optical processes, and different systems including integrated photonic devices. Overall, our findings could pave the way towards highly compact, efficient, tunable nonlinear optical components for both free-space and on-chip photonics.


\section*{Conflicts of interest}
There are no conflicts to declare.

\section*{Data availability}

All relevant data supporting the findings of this study are included in the article and its Electronic Supplementary Information (SI)$^\dag$, which includes: description of methods (fabrication, experiments, and numerical simulations), additional results (SEM, AFM, ellipsometry, SHG power dependence, simulated results for sample B, TE transmission spectra), and data analysis (determination of the $\chi^{(2)}$ tensor elements and $Q$ factors). Additional data and computational files are available on the Fairdata IDA platform at \href{https://doi.org/10.23729/fd-c80a3f6e-dafb-3d98-9872-7000092dcc6f}{https://doi.org/10.23729/fd-c80a3f6e-dafb-3d98-9872-7000092dcc6f}, Ref.~\cite{kolkowski26}. 

\section*{Acknowledgements}

TS, HB, and RK acknowledge support of the Research Council of Finland (Grants No. 347449, 353758 and 368485). TS acknowledges Jenny and Antti Wihuri Foundation for their postdoctoral research grant. RK has received funding from the European Union’s Horizon Europe programme for research and innovation under the Marie Skłodowska-Curie Grant Agreement No.~101060306 (NExIA). The authors acknowledge Sagar Sehrawat, Serguei Novikov, and Juan Camilo Arias for their help in sample fabrication and Biao Chen for his help in numerical simulations. The authors acknowledge the provision of facilities and technical support by Aalto University OtaNano – Nanofabrication Centre for sample fabrication. The calculations were performed using computer resources within the Aalto University School of Science “Science-IT” project.



\balance


\bibliography{rsc} 
\bibliographystyle{rsc} 
\end{document}


\setstretch{1.25}

\,
\vspace{1cm}

\par{\centering\LARGE{\textbf{Supplementary Information for \\Doubly resonant enhancement of second-harmonic generation with in-plane phase matching in plasmonic metasurfaces on an AlInP slab waveguide\\}}}

\vspace{4mm}

\par{\centering Timo Stolt,\textit{$^{a}$} Huayu Bai,\textit{$^{a}$} Seyed Ahmad Shahahmadi,\textit{$^{b}$} Jani Oksanen,\textit{$^{b}$}\\ Andriy Shevchenko,\textit{$^{a\ast}$} and Radoslaw Kolkowski$^{a\ast}$\\}

\vspace{4mm}

\par{\centering\textit{$^{a}$~Department of Applied Physics, Aalto University, P.O.Box 13500, FI-00076 Aalto, Finland}\\
\textit{$^{b}$~Engineered Nanosystems Group, Aalto University, P.O.Box 13500, FI-00076 Aalto, Finland}\\
\textit{$^{*}$~E-mail: andriy.shevchenko@aalto.fi, radoslaw.kolkowski@aalto.fi}\\}

\vspace{1cm}

\setstretch{0.9}

\tableofcontents

\setstretch{1.25}

\newpage
\,
\vspace{1cm}
\section{\hspace{5mm}Sample fabrication\label{ss:I}}\vspace{2mm}
A 400 nm thick AlInP layer was epitaxially grown on a GaAs wafer by metal–organic vapor phase epitaxy (MOVPE). The layer was then transferred onto a quartz substrate via adhesive wafer bonding, followed by selective wet chemical etching~\cite{holmes1995high} (with the help of etch-stop layers) to remove the GaAs substrate. The obtained AlInP layer bonded to quartz was investigated by spectroscopic ellipsometry and atomic force microscopy (see Sections \ref{ss:III} and \ref{ss:IV}). 

To fabricate the metasurfaces, a 200 nm thick PMMA resist was spin-coated on the AlInP surface and patterned by electron-beam lithography (EBL). The pattern was developed in a 1:3 MIBK:IPA solution. Next, a 40 nm thick Au film was deposited by electron-beam evaporation. Finally, a lift-off was performed in acetone for 4 hours without ultrasonic agitation or heating to avoid degradation of the used adhesive. The obtained samples were characterized using scanning electron microscopy (SEM; see Section \ref{ss:II}).

\vspace{1cm}

\section{\hspace{5mm}Scanning electron microscopy images of the samples\label{ss:II}}\vspace{2mm}

\begin{figure}
    \centering
    \includegraphics[width=0.725\textwidth]{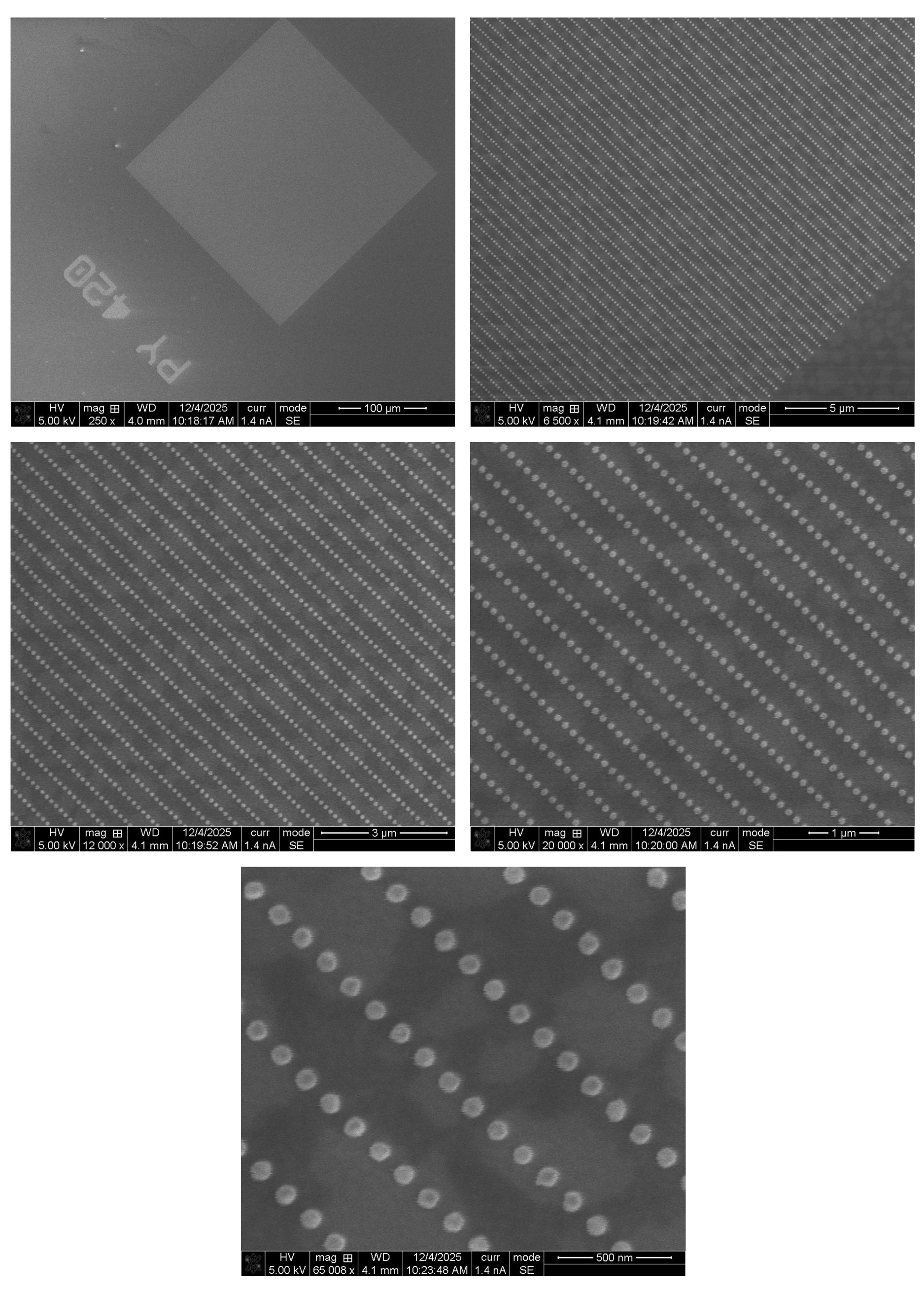}\vspace{-2mm}
    \caption{\centering SEM images of sample A ($P_y$ = 420 nm).}
    \label{fig:SEM_A}
\end{figure}

\begin{figure}
    \centering
    \includegraphics[width=0.725\textwidth]{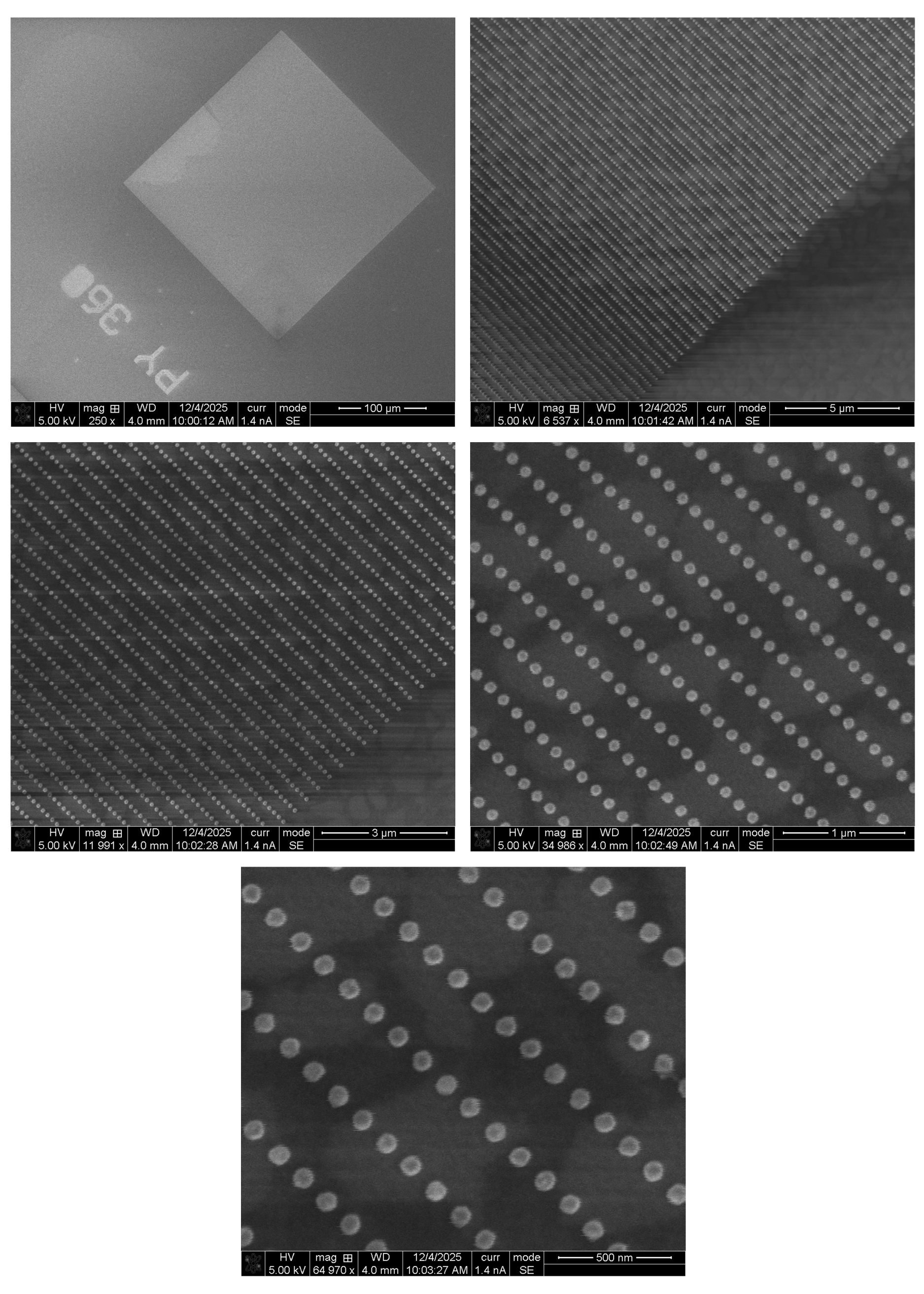}\vspace{-2mm}
    \caption{\centering SEM images of sample B ($P_y$ = 360 nm).}
    \label{fig:SEM_B}
\end{figure}

Figures~\ref{fig:SEM_A} and \ref{fig:SEM_B} show representative SEM images of samples A and B, respectively. The images were obtained at different magnifications, starting from the zoomed-out views of the whole metasurfaces (top left) and ending at the close-ups of the gold nanodiscs (bottom). At intermediate magnifications, the images reveal the high regularity of the lattice periods.

\newpage

\,
\vspace{1cm}
\section{\hspace{5mm}Atomic force microscopy of the AlInP surface\label{ss:III}}\vspace{2mm}

\begin{figure}[h]
    \centering
    \includegraphics[width=0.6\textwidth]{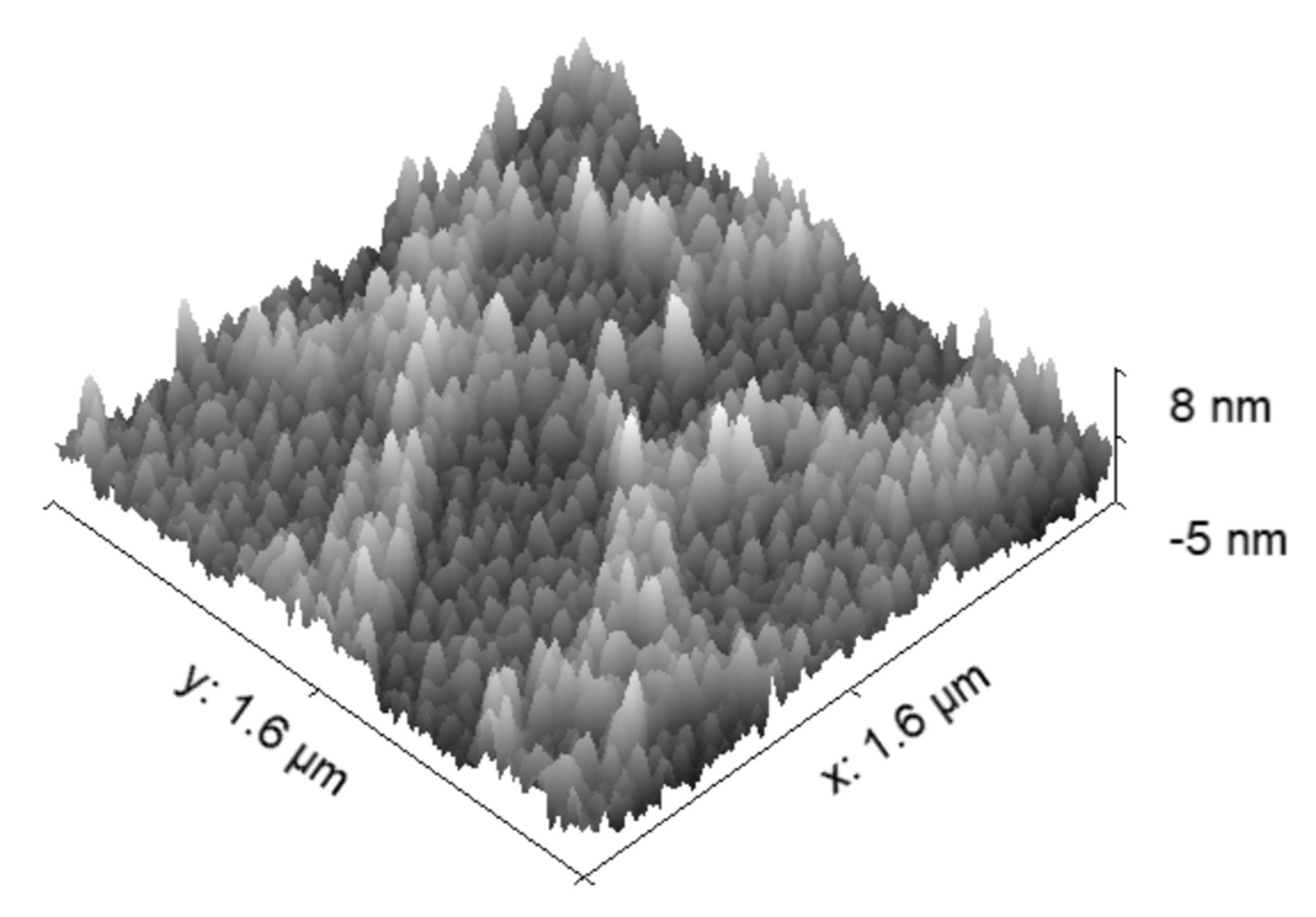}
    \caption{\centering Surface roughness of AlInP measured by AFM. The mean roughness is 1.5 nm. \\The maximum peak height is 8.1 nm and the maximum pit depth is 5.2 nm.}
    \label{fig:afm}
\end{figure}

We characterized the surface roughness of AlInP using atomic force microscopy (AFM). The results are shown in Fig.~\ref{fig:afm}. The AlInP surface exposed by the chemical etching of GaAs appears to be relatively smooth. On the small scale (sub-100 nm in the lateral directions), the topography is dominated by shallow grains (height on the order of 1--2 nm), and on the larger scale ($\sim$1~$\upmu$m), by branched ridges (height 5--10 nm).

\newpage

\,
\vspace{1cm}
\section{\hspace{5mm}Ellipsometry results for the AlInP layer\label{ss:IV}}\vspace{2mm}

\begin{figure}[h]
    \centering
    \includegraphics[width=0.95\textwidth]{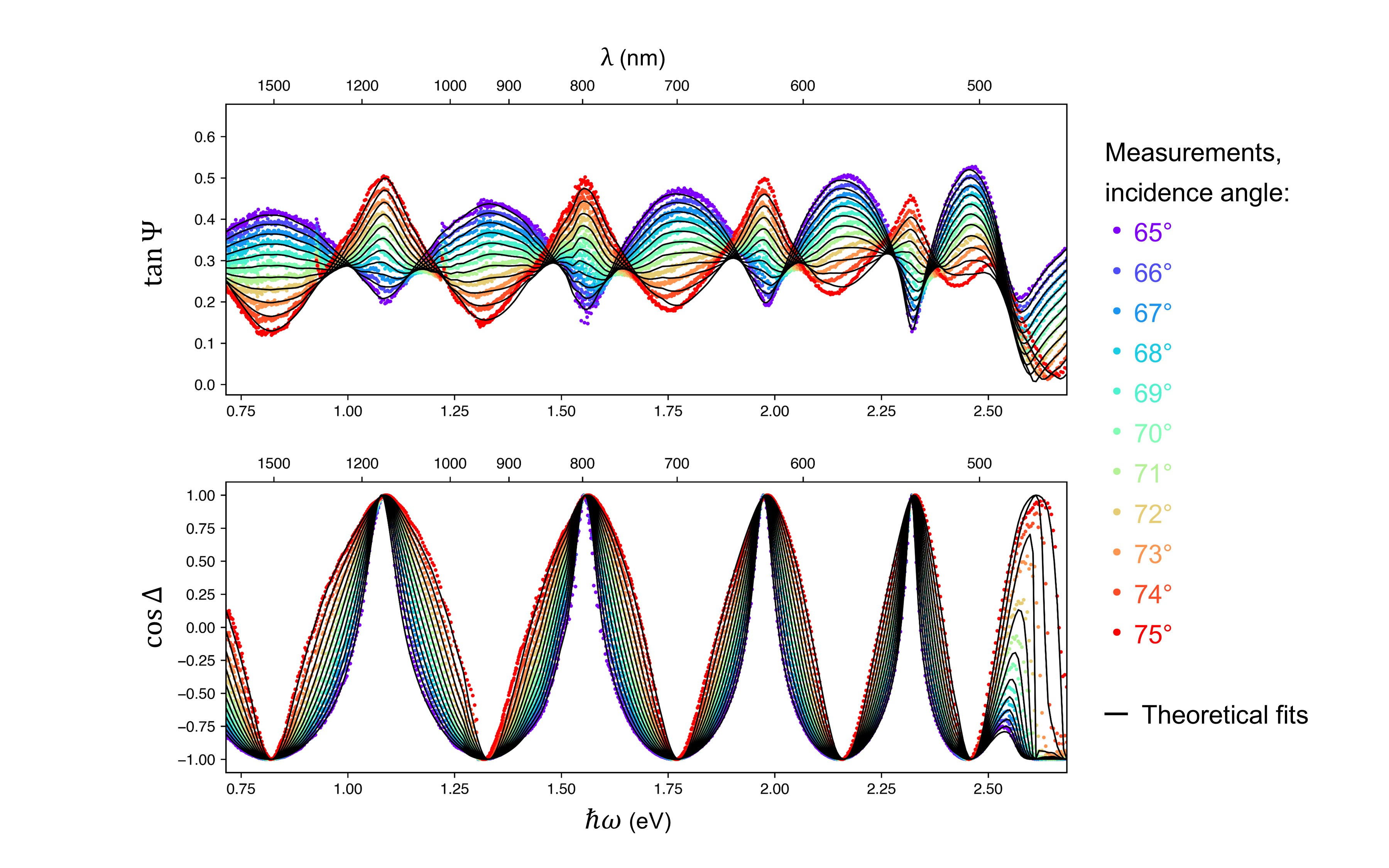}\vspace{-2mm}
    \caption{Ellipsometry data ($\tan{\Psi}$ and $\cos{\Delta}$ in the top and bottom plots, respectively) measured from the AlInP layer bonded to quartz. The colors correspond to different incidence angles, and the black solid lines show the theoretical fits. The fitted AlInP thickness is 406.5 nm. The extracted optical constants are shown in Fig.~\ref{fig:nk} and Table~\ref{tab:t1}.}
    \label{fig:ellipsometry}
\end{figure}

\begin{figure}
    \centering
    \includegraphics[width=0.95\textwidth]{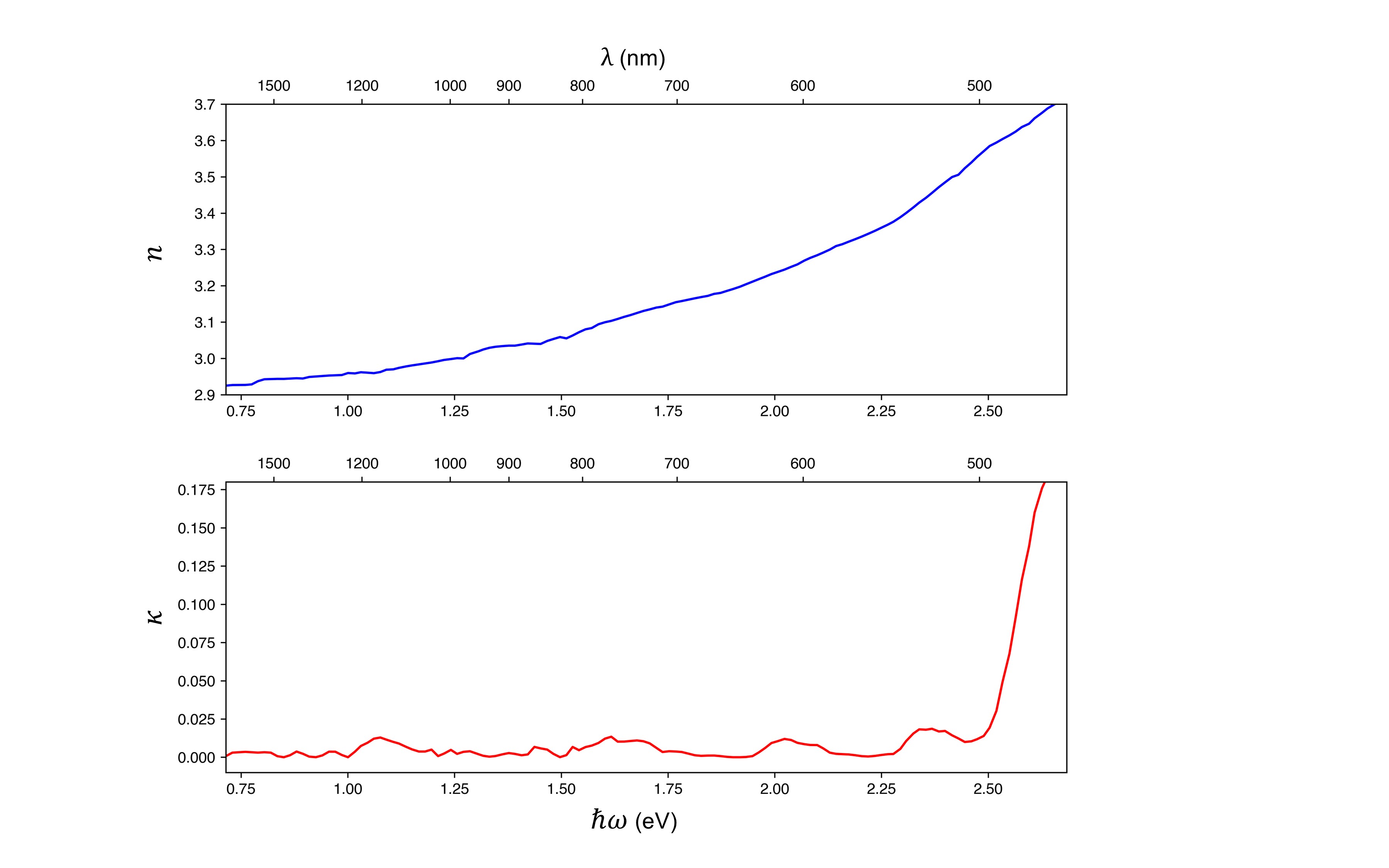}\vspace{-2mm}
    \caption{\centering Real and imaginary parts of the refractive index ($n$ and $\kappa$ in the top and bottom plot, respectively) obtained by fitting the ellipsometry data in Fig.~\ref{fig:ellipsometry}.}
    \label{fig:nk}
\end{figure}

\begin{figure}
    \centering
    \includegraphics[width=0.95\textwidth]{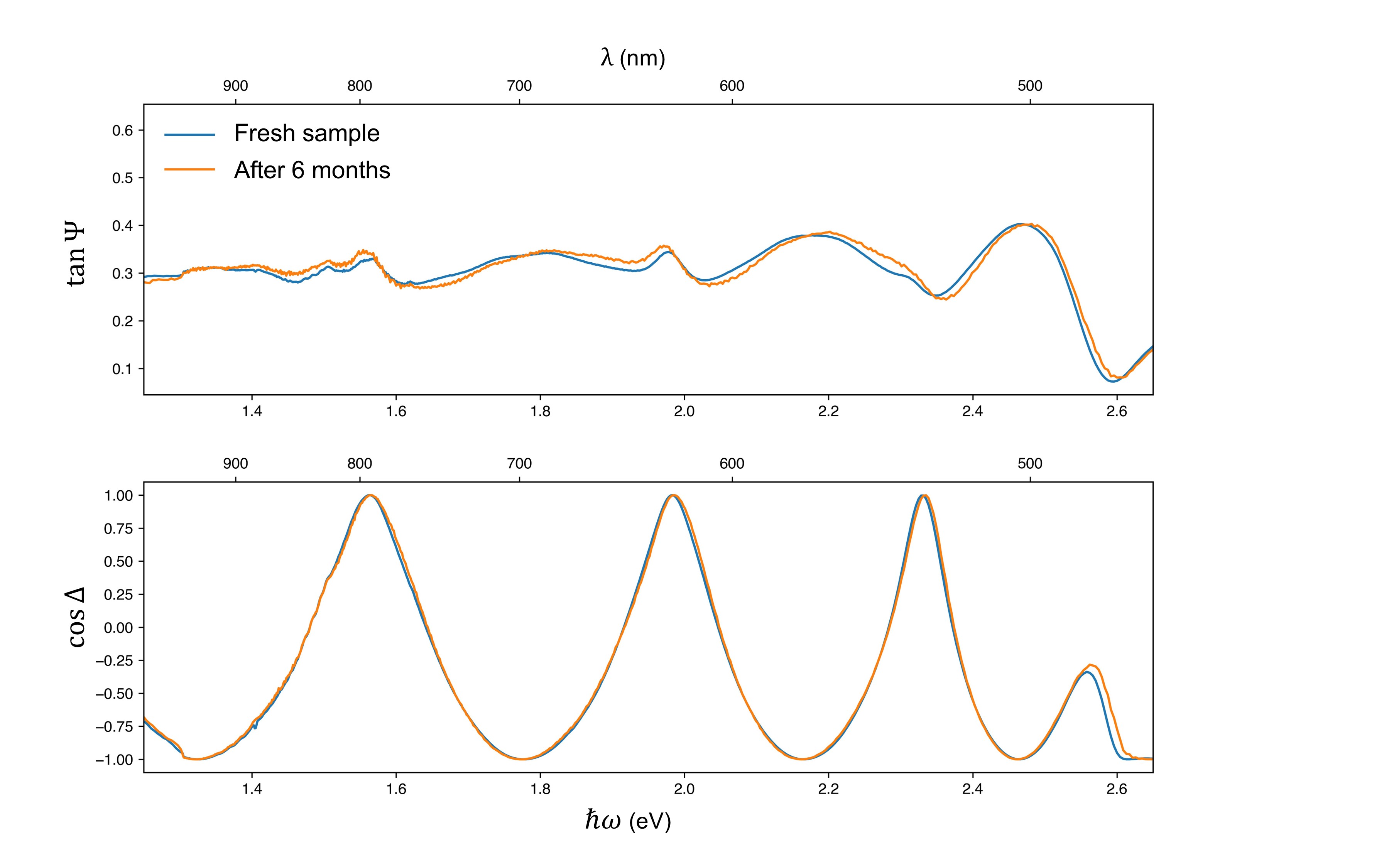}\vspace{-2mm}
    \caption{\centering Ellipsometry data for the AlInP layer at the incidence angle of $\ang{70}$, measured directly after exposing the AlInP surface by wet etching of GaAs (blue curve) and six months later (orange curve).}
    \label{fig:ellipsometry_6months}
\end{figure}

The optical properties of AlInP were studied using spectroscopic ellipsometry at variable incidence angles. The results of these measurements are presented in Fig.~\ref{fig:ellipsometry}. The fitted optical constants ($n$, $\kappa$) are shown in Fig.~\ref{fig:nk} and their values are provided in Table~\ref{tab:t1}. In addition, we studied the stability of the ellipsometric curves over six months. The results are presented in Fig.~\ref{fig:ellipsometry_6months}, showing that slow oxidation of AlInP has a very small effect on its optical properties.

\begin{table}[th!]
\caption{\label{tab:t1} \centering Optical constants of AlInP extracted from the ellipsometry measurements}
\begin{center}
\scriptsize
\begin{tabular}{|ccc|}
$\lambda$ ($\upmu$m) & $n$ & $\kappa$ \\
0.461923 & 3.737529 & 0.203645 \\
0.464276 & 3.709873 & 0.193388 \\
0.467021 & 3.699110 & 0.183131 \\
0.469766 & 3.688347 & 0.184009 \\
0.472119 & 3.676113 & 0.175904 \\
0.475256 & 3.661155 & 0.159931 \\
0.477609 & 3.646198 & 0.138007 \\
0.480746 & 3.637054 & 0.116083 \\
0.483491 & 3.624424 & 0.091186 \\
0.486236 & 3.614155 & 0.067694 \\
0.489373 & 3.603886 & 0.049022 \\
0.492118 & 3.594324 & 0.030350 \\
0.495255 & 3.584763 & 0.019306 \\
0.498000 & 3.570456 & 0.013912 \\
0.501137 & 3.554799 & 0.011789 \\
0.503882 & 3.539284 & 0.010397 \\
0.507019 & 3.523768 & 0.009968 \\
0.510156 & 3.505622 & 0.012287 \\
0.513293 & 3.499384 & 0.014329 \\
0.516822 & 3.484690 & 0.017182 \\
0.519959 & 3.471538 & 0.016837 \\
0.523489 & 3.455633 & 0.018612 \\
0.526626 & 3.442155 & 0.017983 \\
0.530155 & 3.429062 & 0.018233 \\
0.533292 & 3.415788 & 0.015533 \\
0.537213 & 3.400399 & 0.010595 \\
0.540350 & 3.388946 & 0.005501 \\
0.544272 & 3.376260 & 0.002125 \\
0.547409 & 3.368244 & 0.001977 \\
0.551330 & 3.359227 & 0.001427 \\
0.554859 & 3.351144 & 0.000844 \\
0.558781 & 3.342963 & 0.000480 \\
0.562702 & 3.335181 & 0.000714 \\
0.566623 & 3.328028 & 0.001332 \\
0.570545 & 3.321295 & 0.001818 \\
0.574466 & 3.314384 & 0.002006 \\
0.578388 & 3.309321 & 0.002219 \\
0.582309 & 3.299698 & 0.002978 \\
0.586622 & 3.291310 & 0.005802 \\
0.590544 & 3.284106 & 0.007958 \\
0.595249 & 3.276593 & 0.008018 \\
0.599171 & 3.269245 & 0.008541 \\
0.603876 & 3.258807 & 0.009411 \\
0.608190 & 3.252129 & 0.011258 \\
\end{tabular}\qquad
\begin{tabular}{|ccc|}
$\lambda$ ($\upmu$m) & $n$ & $\kappa$ \\
0.612895 & 3.244523 & 0.011967 \\
0.617601 & 3.238277 & 0.010539 \\
0.622307 & 3.232239 & 0.009280 \\
0.627012 & 3.224933 & 0.005994 \\
0.631718 & 3.218164 & 0.003128 \\
0.636424 & 3.211395 & 0.000733 \\
0.641521 & 3.204119 & 0.000148 \\
0.646227 & 3.197399 & 0.000000 \\
0.651717 & 3.191061 & 0.000000 \\
0.656815 & 3.185739 & 0.000260 \\
0.661912 & 3.180324 & 0.000733 \\
0.667402 & 3.177624 & 0.001108 \\
0.672500 & 3.172271 & 0.001070 \\
0.678382 & 3.168865 & 0.000910 \\
0.683872 & 3.165516 & 0.001290 \\
0.689754 & 3.161676 & 0.002364 \\
0.695636 & 3.157927 & 0.003420 \\
0.701126 & 3.154797 & 0.003733 \\
0.707400 & 3.148660 & 0.003914 \\
0.713674 & 3.142524 & 0.003409 \\
0.719556 & 3.140073 & 0.006071 \\
0.726223 & 3.135002 & 0.009060 \\
0.732497 & 3.130750 & 0.010436 \\
0.739163 & 3.125057 & 0.010964 \\
0.745829 & 3.119363 & 0.010614 \\
0.752496 & 3.114553 & 0.010232 \\
0.759554 & 3.108747 & 0.010196 \\
0.766613 & 3.103412 & 0.013403 \\
0.773671 & 3.099659 & 0.012172 \\
0.781122 & 3.094144 & 0.009323 \\
0.788964 & 3.083664 & 0.007608 \\
0.796415 & 3.080124 & 0.006638 \\
0.804258 & 3.072152 & 0.004600 \\
0.812100 & 3.063102 & 0.006707 \\
0.819943 & 3.055240 & 0.001369 \\
0.828178 & 3.058870 & 0.000051 \\
0.836805 & 3.053482 & 0.002204 \\
0.845040 & 3.048094 & 0.004994 \\
0.854059 & 3.039879 & 0.005781 \\
0.862686 & 3.040474 & 0.006755 \\
0.872097 & 3.041266 & 0.001829 \\
0.881116 & 3.038220 & 0.001283 \\
0.890919 & 3.035173 & 0.002164 \\
0.900331 & 3.035211 & 0.002724 \\
\end{tabular}\qquad
\begin{tabular}{|ccc|}
$\lambda$ ($\upmu$m) & $n$ & $\kappa$ \\
0.910526 & 3.033868 & 0.001851 \\
0.921114 & 3.032099 & 0.000780 \\
0.930917 & 3.029303 & 0.000353 \\
0.941191 & 3.024581 & 0.000931 \\
0.948913 & 3.019859 & 0.001921 \\
0.964015 & 3.012183 & 0.003905 \\
0.975550 & 2.999990 & 0.003515 \\
0.987085 & 3.000693 & 0.002188 \\
0.998620 & 2.998258 & 0.004801 \\
1.011803 & 2.995823 & 0.002377 \\
1.023338 & 2.992306 & 0.000753 \\
1.036521 & 2.988790 & 0.004973 \\
1.049703 & 2.986183 & 0.003726 \\
1.062886 & 2.983576 & 0.003729 \\
1.077717 & 2.980782 & 0.005134 \\
1.090900 & 2.977988 & 0.006797 \\
1.107379 & 2.973969 & 0.008979 \\
1.120561 & 2.969950 & 0.010030 \\
1.137040 & 2.968929 & 0.011499 \\
1.151871 & 2.962587 & 0.012887 \\
1.168349 & 2.959484 & 0.012144 \\
1.184828 & 2.960723 & 0.009487 \\
1.202955 & 2.961961 & 0.007356 \\
1.219433 & 2.958728 & 0.003627 \\
1.239208 & 2.959822 & 0.000000 \\
1.257334 & 2.954300 & 0.001459 \\
1.277108 & 2.953597 & 0.003601 \\
1.296883 & 2.952894 & 0.003633 \\
1.318305 & 2.951583 & 0.001191 \\
1.339727 & 2.950289 & 0.000030 \\
1.362797 & 2.948994 & 0.000382 \\
1.385867 & 2.944709 & 0.002278 \\
1.408937 & 2.945474 & 0.003715 \\
1.433655 & 2.944534 & 0.001310 \\
1.458373 & 2.943631 & 0.000000 \\
1.484739 & 2.943711 & 0.000684 \\
1.512753 & 2.943179 & 0.003009 \\
1.540766 & 2.942648 & 0.003262 \\
1.570428 & 2.937282 & 0.003014 \\
1.600089 & 2.928512 & 0.003262 \\
1.631399 & 2.927015 & 0.003509 \\
1.664356 & 2.926889 & 0.003254 \\
1.698961 & 2.926763 & 0.002999 \\
1.735214 & 2.925104 & 0.000732 \\
\end{tabular}
\end{center}
\end{table}

\newpage
\,
\vspace{-1mm}

\section{\hspace{5mm}Experimental setup\label{ss:V}}\vspace{2mm}

\begin{figure}[h]
    \centering\vspace{4mm}
    \includegraphics[width=1\textwidth]{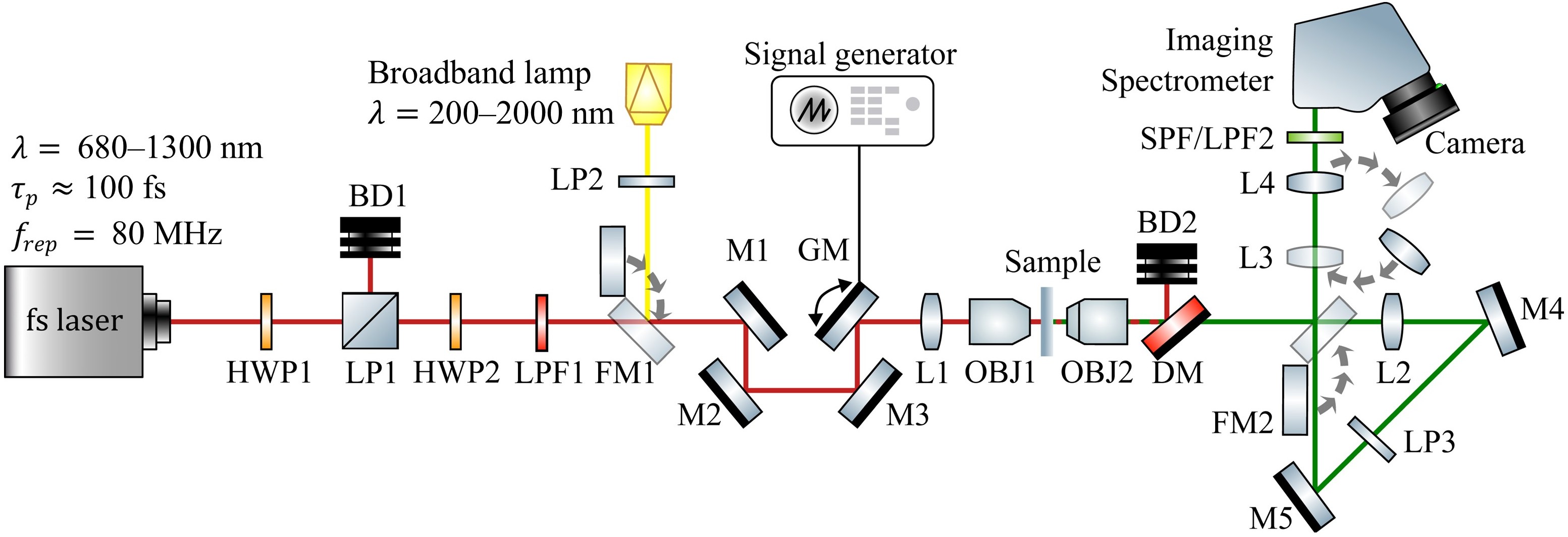}
    \caption{\centering Schematic of the used experimental setup. HWP - half-wave plate, LP - linear polarizer, \\BD - beam dump, LPF - long-pass filter, M - mirror, FM - flip mirror, GM - galvo mirror, L - lens, \\OBJ - objective, DM - dichroic mirror, SPF - short-pass filter.}
    \label{fig:setup}
\end{figure}

The schematic diagram of our experimental setup is presented in Fig.~\ref{fig:setup}. In the SHG measurements, we used a femtosecond laser (Spectra-Physics, InSight DS+) with wavelength tunable in the range of $\lambda=\si{680}$--$\SI{1300}{nm}$, pulse duration $\tau_p$ of about $\SI{100}{fs}$, and repetition rate $f_{rep}=\SI{80}{MHz}$. The power level of the pump beam was controlled  with a half-wave plate (HWP) and a linear polarizer (LP; in this case: a polarizing cube beamsplitter). Another HWP was used to control the polarization, and a long-pass filter (LPF) blocked any unwanted short-wavelength radiation propagating with the pump. 

In the transmission measurements, we used a quartz tungsten-halogen lamp (Thorlabs,\break  QTH10/M), emitting in a broad spectral range of $\lambda=\si{200}$--$\SI{2000}{nm}$. The beam was collimated, spatially filtered, and polarized using another LP. A flip-mirror was used to switch between the two light sources.

The rest of the setup was similar for both the nonlinear and transmission measurements. The light was directed to a galvo mirror (GM), whose tilt was controlled with an applied voltage. The reflected beam was focused by an aspheric lens ($f=\SI{50}{mm}$) in the back-focal plane of an objective (OBJ1, OptoSigma, PAL-20-NIR-HR-LC00, NA $=0.45$). The sample was mounted on a manual translation stage combined with a goniometer and positioned in the focus of OBJ1. Such a configuration transfers the GM tilt into the incidence angle $\theta_y$ on the sample. To scan the incidence angle, we  used a periodic ramp-type electric signal from a function generator to drive GM. The transmitted light (or the SHG emission from the sample) was collected by another objective (OBJ2, Zeiss, N-Achroplan 63$\times$, NA $=0.95$) and imaged on the entrance slit of the imaging spectrometer (HORIBA, iHR320) with a sCMOS camera (Hamamatsu, ORCA-Quest). The last filter before the slit was either a short-pass filter (SPF) for measuring the SHG signal in the visible wavelength range, or another long-pass filter to measure the transmission in the infrared range. For transmission measurements in the visible range, all filters were removed. In the imaging part of the setup, we switched between two different configurations of lenses and mirrors (see the scheme in Fig.~\ref{fig:setup}) to perform either real-space or $k$-space imaging. At the same time, by rotating the grating inside the spectrometer (selecting either the 0$^{\text{th}}$ or the 1$^{\text{st}}$ diffraction order) and modifying the width of the entrance slit, we switched between the wide-field and spectrally resolved imaging.

\newpage
\,
\vspace{1cm}

\section{\hspace{5mm}Power dependence of the SHG signal\label{ss:VI}}
\begin{figure}[h]
    \centering\vspace{4mm}
    \includegraphics[width=0.45\textwidth]{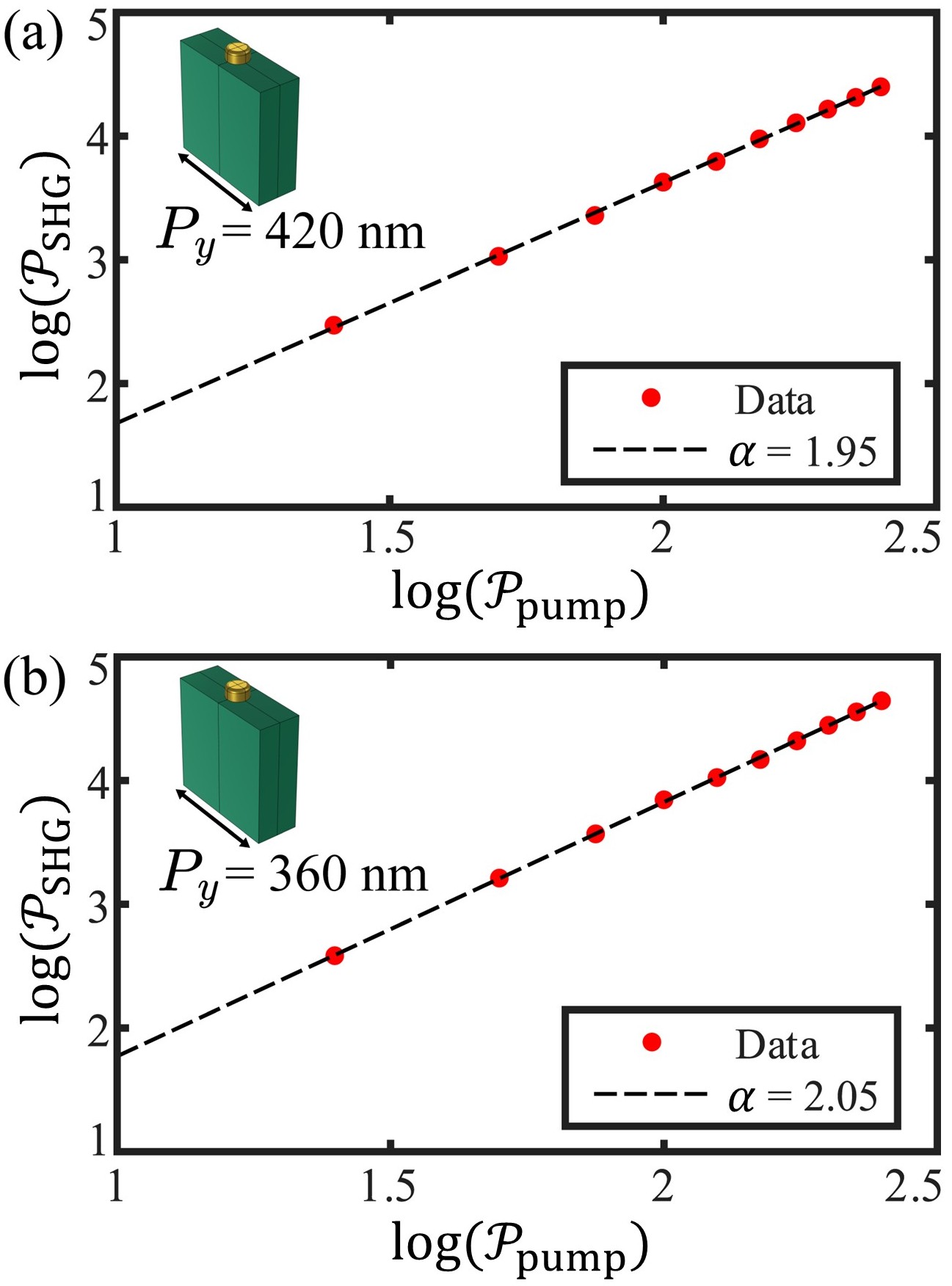}
    \caption{Dependence of the measured SHG signal $\mathcal{P}_{\text{SHG}}$ on the average pump power $\mathcal{P}_{\text{pump}}$, plotted on a double-logarithmic scale, for samples A ($P_y=\SI{420}{nm}$) and B ($P_y=\SI{360}{nm}$) in panels (a) and (b), respectively. The red dots show the measured data points, while the dashed black lines show the linear fit to the data, with $\alpha$ being the slope of the fit. The units of $\mathcal{P}_{\text{SHG}}$ and $\mathcal{P}_{\text{pump}}$ are counts/s and mW, respectively.}\vspace{1cm}
    \label{fig:SHG_vs_P}
\end{figure}

\vspace{2mm}
For both samples, we measured the SHG signal as a function of the average pump power. The measured data are plotted on a log-log scale in  Fig.~\ref{fig:SHG_vs_P}. We fitted these plots with a linear function $\log(\mathcal{P}_{\text{SHG}})\propto \alpha\log(\mathcal{P}_{\text{pump}})$. In both cases, the fitted slope is $\alpha\approx2$, indicating a quadratic power dependence, which is a key characteristic of SHG. Furthermore, the dependence did not show saturation up to $\mathcal{P}_{\text{pump}}=\SI{250}{mW}$, which was the highest power used in the experiments.

\newpage

\section{\hspace{5mm}Data and labels from Fig.~3a-b shown separately\label{ss:VII}}\vspace{2mm}

\begin{figure}[h]
    \centering
    \includegraphics[width=\textwidth]{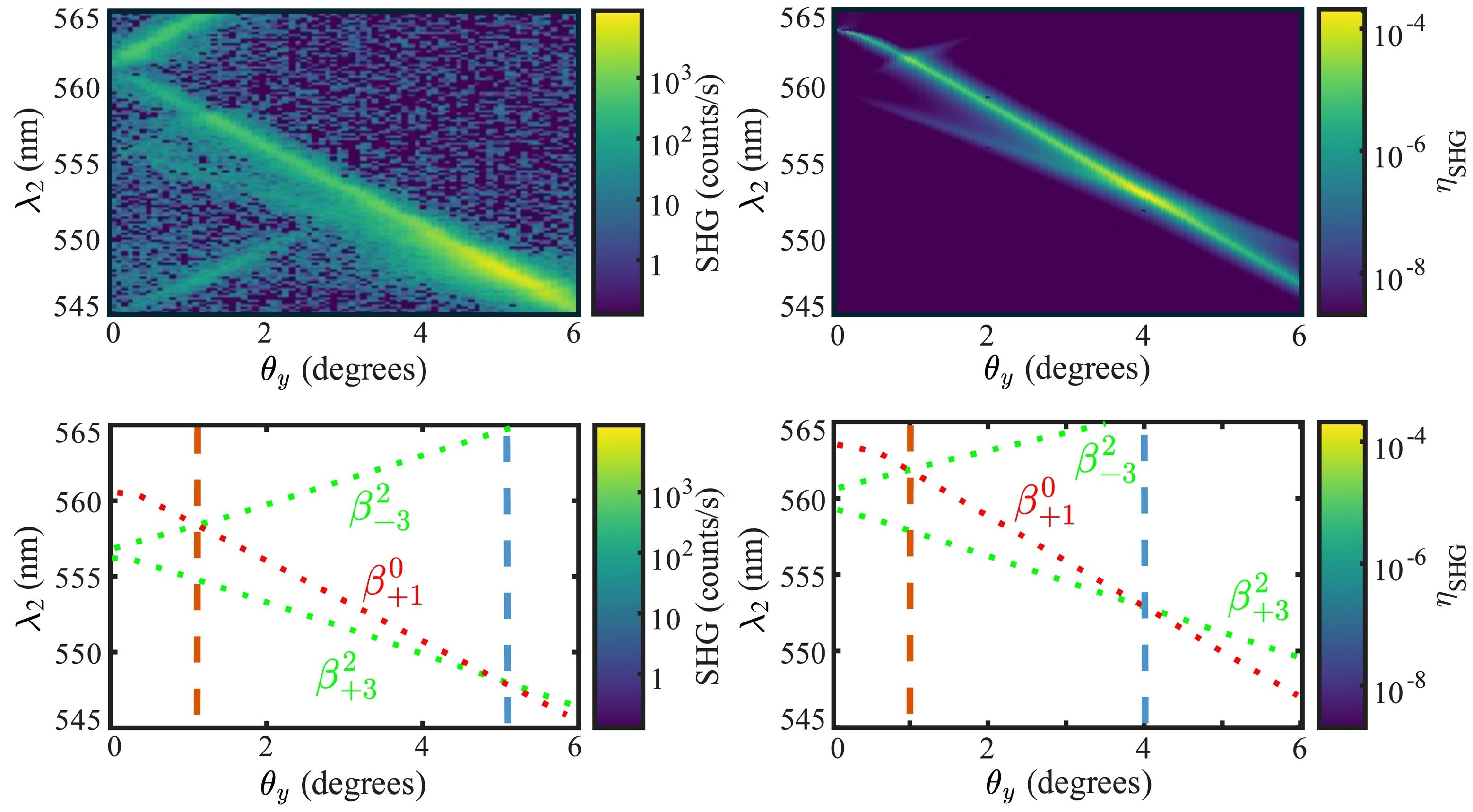}
    
    \vspace{2mm}\caption{\centering The same data and labels as in Fig.~3a and b, but shown in separate panels.}\label{fig:S_3}
\end{figure}

\vspace{1cm}

\section{\hspace{5mm} Extracting the $Q$ factor from the SHG map\label{ss:VIII}}\vspace{2mm}

\begin{figure}[h!]
    \centering
    \includegraphics[width=0.9\textwidth]{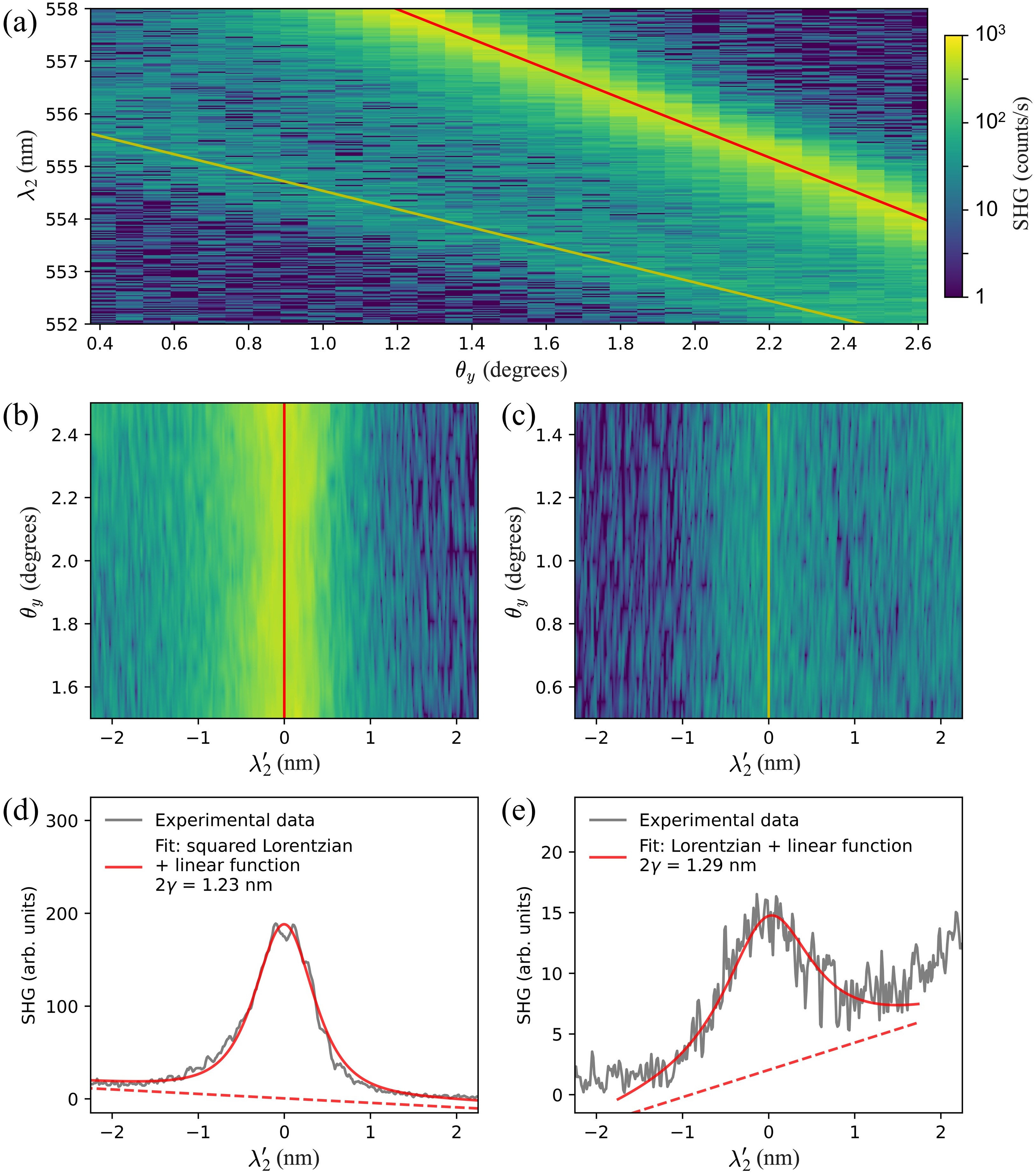}\caption{Illustration of the procedure to extract the GMR's $Q$ factors from the measured dependence of the SHG on $\lambda_2$ and $\theta_y$. (a) Part of the original data (the same as in Figs.~2c and 3a), with the GMR dispersion lines fitted with linear functions (red and yellow lines for the $\beta_{+1}^0$ and $\beta_{+3}^2$ modes, respectively). (b, c) The same data as in (a), but compensated for the spectral shifts of the GMR peaks and linearly interpolated over a dense regular grid. The values displayed in the plots correspond to the color bar shown in (a); note the logarithmic scale. (d, e) Averaged GMR peaks ($\beta_{+1}^0$ and $\beta_{+3}^2$ modes, respectively) obtained by integrating the data in (b, c) along the $\theta_y$-axis. The $Q$ factors are extracted by fitting the data with functions $C/(1+(\lambda'_2)^2/\gamma^2)^2+A\lambda'_2+B$ and $C/(1+(\lambda'_2)^2/\gamma^2)+A\lambda'_2+B$, which yields the spectral widths of the peaks $\Delta\lambda=2\gamma$. We obtained $Q=2\lambda_2/\Delta\lambda\approx900$ for the $\beta_{+1}^0$ mode and $Q=\lambda_2/\Delta\lambda\approx430$ for the $\beta_{+3}^2$ mode. }\label{fig:S_Qfactor}
\end{figure}

Figure~\ref{fig:S_Qfactor} demonstrates the procedure that we used to extract the $Q$ factors from the measured SHG powers as functions of $\lambda_2$ and $\theta_y$. First, we compensated for the linear angular dispersion of the GMRs -- we fitted the data with functions $\lambda_2=A\theta_y+B$ (see the red and yellow lines in Fig.~\ref{fig:S_Qfactor}a) using the magnitude of the SHG signal as weights. Next, we converted the wavelength axis to $\lambda'_2=\lambda_2-(A\theta_y+B)$ and linearly interpolated the data over a dense regular grid in the $\lambda'_2$-$\theta_y$ parameter space (see Fig.~\ref{fig:S_Qfactor}b,c). Finally, we integrated the obtained 2D maps along the $\theta_y$-axis to get the averaged GMR lineshapes, which we fitted with modified Lorentzian functions (see Fig.~\ref{fig:S_Qfactor}d,e). In the case of the first peak, corresponding to the pump-resonant $\beta_{+1}^0$ mode, we fitted the data using the sum of  a squared Lorentzian curve and a linear function, $C/(1+(\lambda'_2)^2/\gamma^2)^2+A\lambda'_2+B$, while for the second peak, i.e., the second-harmonic mode $\beta_{+3}^2$, we used a similar function but without squaring the Lorentzian curve, $C/(1+(\lambda'_2)^2/\gamma^2)+A\lambda'_2+B$. The fitted parameter $2\gamma$ yielded the full width at half maximum ($\Delta\lambda=2\gamma$), which we used to calculate the $Q$ factor from $Q=\lambda_1/\Delta\lambda=2\lambda_2/\Delta\lambda$ for the $\beta_{+1}^0$ mode and $Q=\lambda_2/\Delta\lambda$ for the $\beta_{+3}^2$ mode.

\newpage

\newpage
\,
\vspace{1cm}
\section{\hspace{5mm}Numerical simulations}\vspace{2mm}

In our numerical simulations, we used the commercial finite-element method software COMSOL Multiphysics. The metasurfaces were modeled as a single unit cell with periodic boundary conditions in the lateral directions ($x$ and $y$), in which we used Floquet wavevector to model the optical response at oblique angles. Along the $z$-direction, the geometry was terminated with a scattering boundary condition  emulating the far-field radiation condition. The metasurface geometry contained the 410 nm thick AlInP layer on a SiO$_2$ substrate, with disc-shaped gold nanoparticles on the top surface (diameter 80 nm, height 40 nm). We used the optical constants of gold and SiO$_2$ from Refs.~\cite{polyanskiy2024,johnson_christy_1972,malitson1965interspecimen}, whereas the optical constants of AlInP were based on our ellipsometric data (see Section \ref{ss:IV}).

The simulations were performed in the frequency domain. To obtain the band diagrams in Fig.~1c-d, and the electric field distributions in Fig.~1e-f, we used the eigenfrequency analysis. To simulate SHG (Fig.~3b, d, and e), we performed scattered-field simulations in two steps, first at the fundamental wavelength and second at the SHG wavelength~\cite{bachelier2010origin,kolkowski2015octupolar,kolkowski2016non}. In the first step, the metasurface was illuminated by a plane wave incident from the substrate side at an angle corresponding to the free-space incidence angle $\theta_y$. The calculated electric field distribution was used to evaluate the nonlinear polarization density. In the second step, the nonlinear polarization density was used as a source of the SH field, and the resulting SHG emission from the metasurface into the far-field was calculated through full-wave numerical simulation performed at the SH frequency. We assumed that the second-order susceptibility is non-zero only in the AlInP layer, with the magnitude of $d_{36}=\SI{39}{pm/V}$~\cite{Shoji2002} and the $\chi^{(2)}$-tensor elements corresponding to the zinc-blende crystal lattice after appropriate transformation (see Section \ref{ss:X}). The SHG intensity $I(\lambda_2)$ was evaluated at a distance of 1 $\upmu$m from the metasurface. The conversion efficiency was obtained from
\begin{equation}
    \eta_{\mathrm{SHG}}=\frac{I(\lambda_2)}{I(\lambda_1)},
\end{equation}
with the incident fundamental wave intensity $I(\lambda_1)$ of $\SI{38.8}{MW/cm^2}$ corresponding to the experimental conditions.

\newpage
\,
\vspace{1cm}

\section{\hspace{5mm}AlInP crystal orientation and the $\chi^{(2)}$ tensor}\vspace{2mm}\label{ss:X}

Similar to many III-V semiconductors, AlInP has a zinc-blende-type crystal structure, belonging to the $\bar{4}3m$ point group. In the conventional lattice orientation (with the [100], [010], and [001] crystallographic directions aligned with the $x$, $y$, and $z$ directions), and assuming that Kleinman symmetry holds, the $\chi^{(2)}$ tensor has only one independent nonzero element with $\chi^{(2)}_{xyz}=\chi^{(2)}_{yxz}=\chi^{(2)}_{zxy}=\chi^{(2)}_{zyx}=\chi^{(2)}_{xzy}=\chi^{(2)}_{yzx}=2d_{36}$~\cite{BoydBook2020}. The nonlinear polarization is then given by

\begin{equation}
    \vb{P}^{(2)}=2\epsilon_0d_{36}\mqty[2E_{y}E_{z} \\ 2E_{x}E_{z} \\ 2E_{x}E_{y}],
    \label{eq:P2_Td}
\end{equation}
where $E_{i}$ are the Cartesian components of the electric field vector.

In our samples, the AlInP crystal lattice is rotated by $\ang{45}$ in the $xy$-plane with respect to the $xyz$ frame of the metasurface and experimental setup. The rotation matrix is
\begin{equation}
    \vb{R}=\mqty[\cos{\ang{45}} & -\sin{\ang{45}} & 0 \\ \sin{\ang{45}} & \cos{\ang{45}} & 0 \\ 0 & 0 & 1].
    \label{eq:P2_Td}
\end{equation}
The $\chi^{(2)}$ tensor should be transformed according to~\cite{jatirian2016calculation}
\begin{equation}
    \chi^{(2)}_{ijk}=\sum_{I=1}^3\sum_{J=1}^3\sum_{K=1}^3\chi^{(2)}_{IJK}R_{iI}R_{jJ}R_{kK},
    \label{eq:P2_Td}
\end{equation}
which gives $\chi^{(2)}_{yyz}=\chi^{(2)}_{yzy}=\chi^{(2)}_{zyy}=-\chi^{(2)}_{xxz}=-\chi^{(2)}_{xzx}=-\chi^{(2)}_{zxx}$. Hence, the nonlinear polarization in AlInP is
\begin{equation}
    \vb{P}^{(2)}=2\epsilon_0d_{36}\mqty[-2E_xE_z\\ 
   2E_yE_z\\
    -E_x^2+E_y^2].
    \label{eq:P2_xyz_45}
\end{equation}
In our experiments, the pump is always $y$-polarized and the incidence angle $\theta_y$ lies in the $yz$-plane. Ignoring the effect of near-field depolarization due to the presence of plasmonic nanoparticles, the measured SHG signal can be assumed to originate mainly from the $\chi^{(2)}_{yyz}$ and $\chi^{(2)}_{zyy}$ elements. Since both the pump and SHG fields contain mainly the $y$- and $z$-components, they are efficiently coupled to the TM-polarized GMRs.

\newpage
\,
\vspace{1cm}
\section{\hspace{5mm}Simulated SHG response for sample B}\vspace{2mm}
\begin{figure}[h]
    \centering
    \includegraphics[width=1\textwidth]{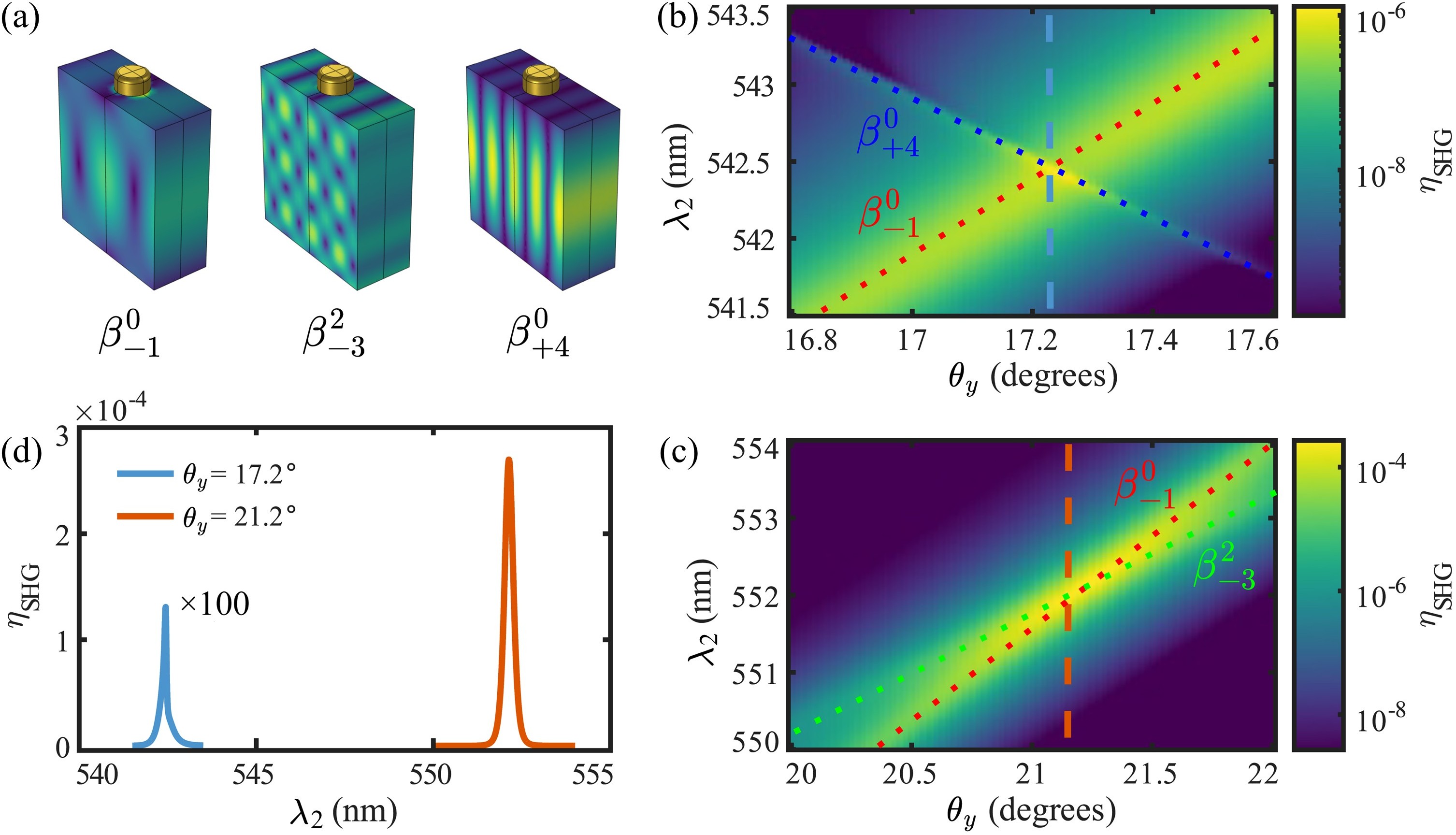}
    \caption{Simulated SHG response for sample B ($P_y=\SI{360}{nm}$). (a) Electric field distributions ($\mathbf{|\mathbf{E}|}$) of the GMRs that contribute to the SHG enhancement, calculated at $\theta_y=0$ using the eigenfrequency analysis in COMSOL. (b, c) Simulated SHG conversion efficiency ($\eta_{\text{SHG}}$) at the doubly resonant conditions as a function of the incidence angle ($\theta_y$) and SHG wavelength ($\lambda_2$). (d) Vertical crosscuts of the plots in (b) and (c) (blue and orange curves, respectively) comparing the peaks at the doubly resonant SHG enhancement. The peak at $\theta_y=\ang{17.2}$ (blue) is multiplied by a factor of 100 to make it visible. The maximum SHG enhancement at $\theta_y=\ang{21.2}$ (orange) is 200 times greater than the maximum enhancement at $\theta_y=\ang{17.2}$.}
    \label{fig:SHG_360_sim}
\end{figure}

\noindent Figure~\ref{fig:SHG_360_sim} shows the results of the SHG simulations for sample B ($P_y=\SI{360}{nm}$). The simulated responses agree well with the experimental results (see Fig.~4 in the main article). In particular, the doubly resonant conditions occur at nearly the same parameter values as in the experiment. Furthermore, the SHG enhancement follows exactly the same trend, becoming significantly stronger for the co-propagating fundamental and SH GMRs.

\newpage
\,
\vspace{1cm}

\section{\hspace{5mm}TE-polarized guided-mode resonances}\vspace{2mm}

\begin{figure}[h]
    \centering
    \includegraphics[width=0.5\textwidth]{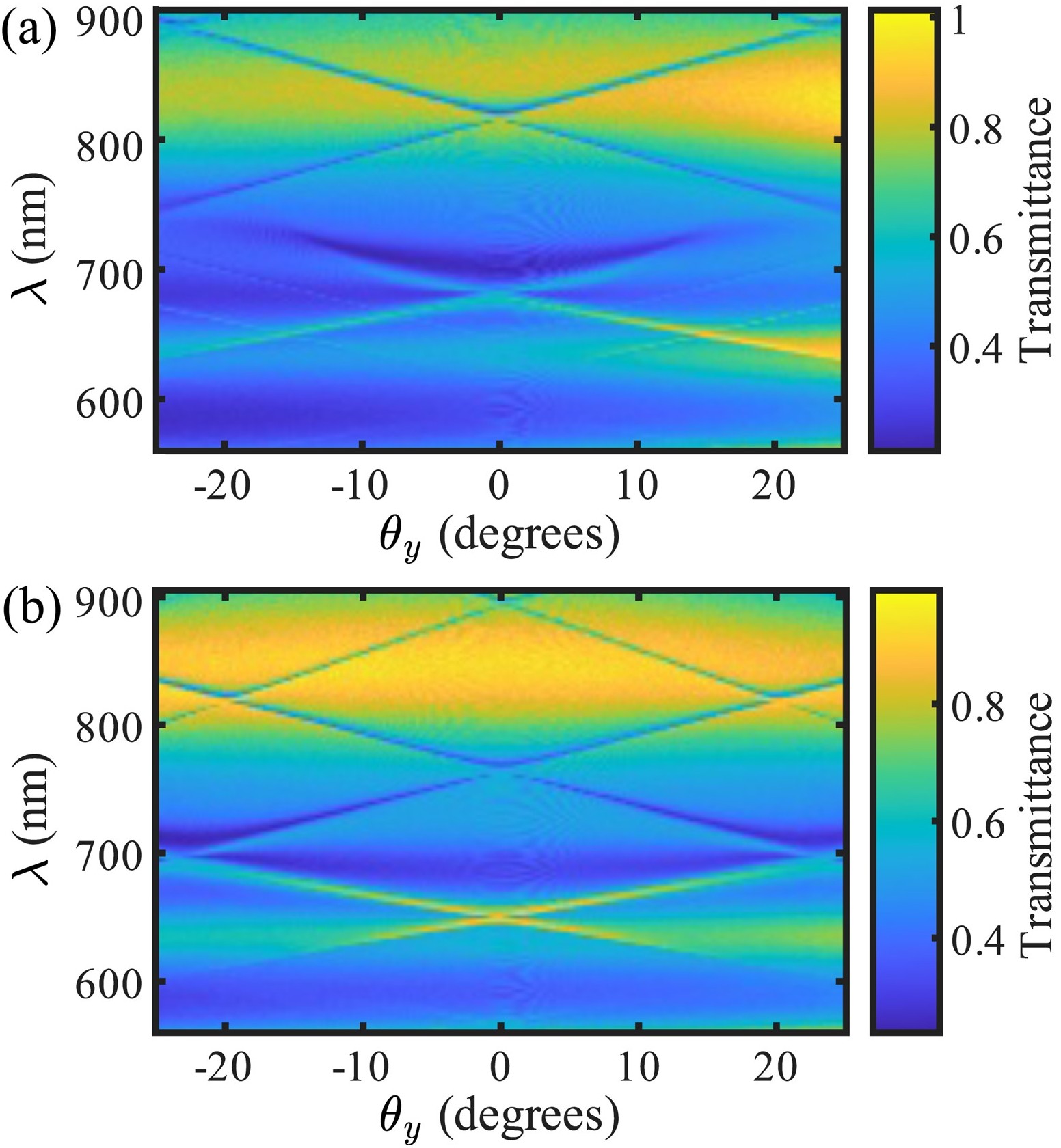}
    \caption{Transmission of broadband $x$-polarized light measured as a function of the wavelength and incidence angle for (a) sample A ($P_y=\SI{420}{nm}$) and (b) sample B ($P_y=\SI{360}{nm}$), revealing the dispersive high-$Q$ TE-polarized GMRs.}
    \label{fig:T_Py360_xpol}
\end{figure}

\noindent In this work, we focused on the TM-polarized GMRs excited by the $y$-polarized incident light, because only these modes have the non-zero electric field components ($E_y$ and $E_z$) that can couple to the non-zero elements of the $\chi^{(2)}$ tensor. However, the studied samples also exhibit transverse electric (TE) GMRs. Although they do not contribute to the SHG enhancement, they give rise to narrow features in the transmission spectra measured under illumination by the $x$-polarized light, as shown in Fig.~\ref{fig:T_Py360_xpol}. Based on their spectral widths, the $Q$ factors can be estimated at about 10$^3$, similar to the TM GMRs.

\newpage
\,
\vspace{1cm}

\bibliography{rsc} 
\bibliographystyle{rsc} 